\documentclass[floatfix,groupedaddress,superscriptaddress,aps,showpacs,amsmath,amssymb,floatfix, prl, longbibliography, twocolumn,a4]{revtex4-2}

\usepackage{graphicx,float}
\usepackage{subfigure}
\usepackage{dcolumn}
\usepackage{bm}
\usepackage{mathrsfs}
\usepackage{txfonts}
\usepackage{CJK}
\usepackage[amsmath]{SIunits}
\usepackage{epsfig}
\usepackage{epstopdf}
\usepackage{lipsum}
\usepackage{color}
\usepackage{placeins}

\usepackage[colorlinks,
linkcolor=blue,
anchorcolor=blue,
citecolor=blue,
filecolor=blue,
menucolor=blue,
runcolor=blue,
urlcolor=blue,
frenchlinks=blue]{hyperref}

\allowdisplaybreaks[2]

\newenvironment{SChinese}{%
\CJKfamily{gbsn}%
\CJKtilde
\CJKnospace}{}

\begin{document}

 \begin{CJK}{UTF8}{}
 \begin{SChinese}

 \title{Quantum Squeezing Induced Optical Nonreciprocity}

 \author{Lei Tang}
 \affiliation{College of Engineering and Applied Sciences, National Laboratory of Solid State Microstructures, and  Collaborative Innovation Center of Advanced Microstructures, Nanjing University, Nanjing 210093, China}

 \author{Jiangshan Tang}
 \affiliation{College of Engineering and Applied Sciences, National Laboratory of Solid State Microstructures, and  Collaborative Innovation Center of Advanced Microstructures, Nanjing University, Nanjing 210093, China}

 \author{Mingyuan Chen}
 \affiliation{College of Engineering and Applied Sciences, National Laboratory of Solid State Microstructures, and  Collaborative Innovation Center of Advanced Microstructures, Nanjing University, Nanjing 210093, China}

 \author{Franco Nori}
 \affiliation{RIKEN Quantum Computing Center, RIKEN Cluster for Pioneering Research, Wako-shi, Saitama 351-0198, Japan}
 \affiliation{Physics Department, The University of Michigan, Ann Arbor, Michigan 48109-1040, USA}

 \author{Min Xiao}
 \affiliation{College of Engineering and Applied Sciences, National Laboratory of Solid State Microstructures, and  Collaborative Innovation Center of Advanced Microstructures, Nanjing University, Nanjing 210093, China}
 \affiliation{Department of Physics, University of Arkansas, Fayetteville, Arkansas 72701, USA}

 \author{Keyu Xia}
 \email{keyu.xia@nju.edu.cn}
 \affiliation{College of Engineering and Applied Sciences, National Laboratory of Solid State Microstructures, and  Collaborative Innovation Center of Advanced Microstructures, Nanjing University, Nanjing 210093, China}
 \affiliation{Jiangsu Key Laboratory of Artificial Functional Materials, Nanjing University, Nanjing 210023, China}

 \date{\today}

 \begin{abstract}
  We propose an all-optical approach to achieve optical nonreciprocity on a chip by quantum squeezing one of two coupled resonator modes. By parametric pumping a nonlinear resonator unidirectionally with \emph{a classical coherent field}, we squeeze the resonator mode in a selective direction due to the phase-matching condition, and induce a chiral photon interaction between two resonators. Based on this chiral interresonator coupling, we achieve an all-optical diode and a three-port quasi-circulator. By applying \emph{a second  squeezed-vacuum field} to the squeezed resonator mode, our nonreciprocal device also works for single-photon pulses. We obtain an isolation ratio of $>40~\deci\bel$ for the diode and fidelity of $>98\%$ for the quasi-circulator, and insertion loss of $<1~\deci\bel$ for both.
We also show that nonreciprocal transmission of strong light can be switched on and off by a relative weak pump light. This achievement implies a nonreciprocal optical transistor.  Our protocol opens up a new route to achieve integrable all-optical nonreciprocal devices permitting chip-compatible optical isolation and nonreciporcal quantum information processing.
 \end{abstract}

 \maketitle

 \end{SChinese}
 \end{CJK}

 Optical nonreciprocal devices, such as optical diodes and circulators, can separate backscattering signals from a light source.
 %
 The conventional magneto-optical approach to achieve optical nonreciprocity (ONR) is difficult to integrate on a chip because it requires strong magnetic fields and bulky materials.
 Developing a new mechanism for magnetic-free ONR is of interest in fundamental physics and promises important applications for on-chip light manipulation.
 Various magnetic-free optical nonreciprocal devices have been theoretically proposed and experimentally demonstrated by exploiting optical nonlinearities~\cite{science.335.447.2012,nat.phys.10.394.2014,nat.photonics.8.524.2014,nat.photon.9.2015,nat.commun.7.13657.2016,prl.120.203904.2018,optica.5.279.2018,nat.photonics.14.369.2020,photonres.9.1218.2021}, spatio-temporal modulation of the medium~\cite{prl.109.033901.2012,nat.phys.10.923.2014,nat.photonics.11.774.2017}, spin-momentum locking in chiral quantum optical systems~\cite{pra.90.2014,science.348.2015,phy.rep.592.2015,nat.photonics.9.796.2015,science.354.2016,nat.541.473.2017,NC.10.580.2019,pra.99.043833.2019,NC.12.3746.2021}, directional optomechanical coupling~\cite{prl.102.213903.2009,nat.photonics.10.657.2016,NC.1797.2018}, moving atomic lattices~\cite{prl.110.093901.2013, prl.110.223602.2013, prl.113.123004.2014}, atomic reservoir engineering~\cite{prl.126.223603.2021}, the Sagnac effect in spinning resonators~\cite{nat.558.569.2018,prl.121.153601.2018,prl.125.143605.2020}, and susceptiblity-momentum locking in atomic gases~\cite{prl.121.203602.2018, prl.121.203602.2018,nat.photonics.12.2018,prapplied.12.054004.2019,prl.123.233604.2019,prl.125.123901.2020,sci.adv.7.eabe8924.2021,prapplied.16.014046.2021}. Optical nonreciprocal devices based on the Kerr nonlinearity of materials are compatible with a chip, but subject to dynamic reciprocity~\cite{nat.photon.9.2015}. Despite many efforts, it is challenging to realize a chip-compatible all-optical nonreciprocal device without moving parts or spatio-temporal modulation.
By coherently driving a $\chi^{(2)}$ microring resonator, one can directionally amplify the single-photon interaction in the resonator via the mode mean field. This mechanism has been exploited to induce chiral normal mode splitting (NMS) and construct on-chip optical isolation~\cite{prl.117.2016}. Nevertheless, the demanded three-mode phase matching in one resonator is a significant challenge, as pointed out in Ref.~\cite{prl.126.133601.2021}.

 Resonator mode squeezing via parametric driving of a $\chi^{(2)}$ resonator has been explored to exponentially amplify the interaction between quantum objects~\cite{prl.114.093602.2015,prl.120.093601.2018,prl.120.093602.2018,prl.122.030501.2019,sci.china.63.2019,pra.100.062501.2019,prl.124.073602.2020,PhysRevLett.126.023602,nat.phys.17.898.2021, PhysRevLett.127.093602}. However, it has not been used to achieve ONR.
Here, we show that high-performance ONR can be achieved by directionally squeezing the resonator mode with a \emph{coherent} laser field. With this chiral quantum squeezing, we achieve an optical diode, a quasi-circulator and, for the first time, a nonreciprocal optical transistor.
 Note that our method is based on directional quantum squeezing and thus conceptually differs from Ref.~\cite{prl.117.2016}. Our method only needs two-mode matching in one resonator and thus greatly simplifies its experimental implementation.

 \begin{figure}[ht]
  \centering
  \includegraphics[width=0.99\linewidth]{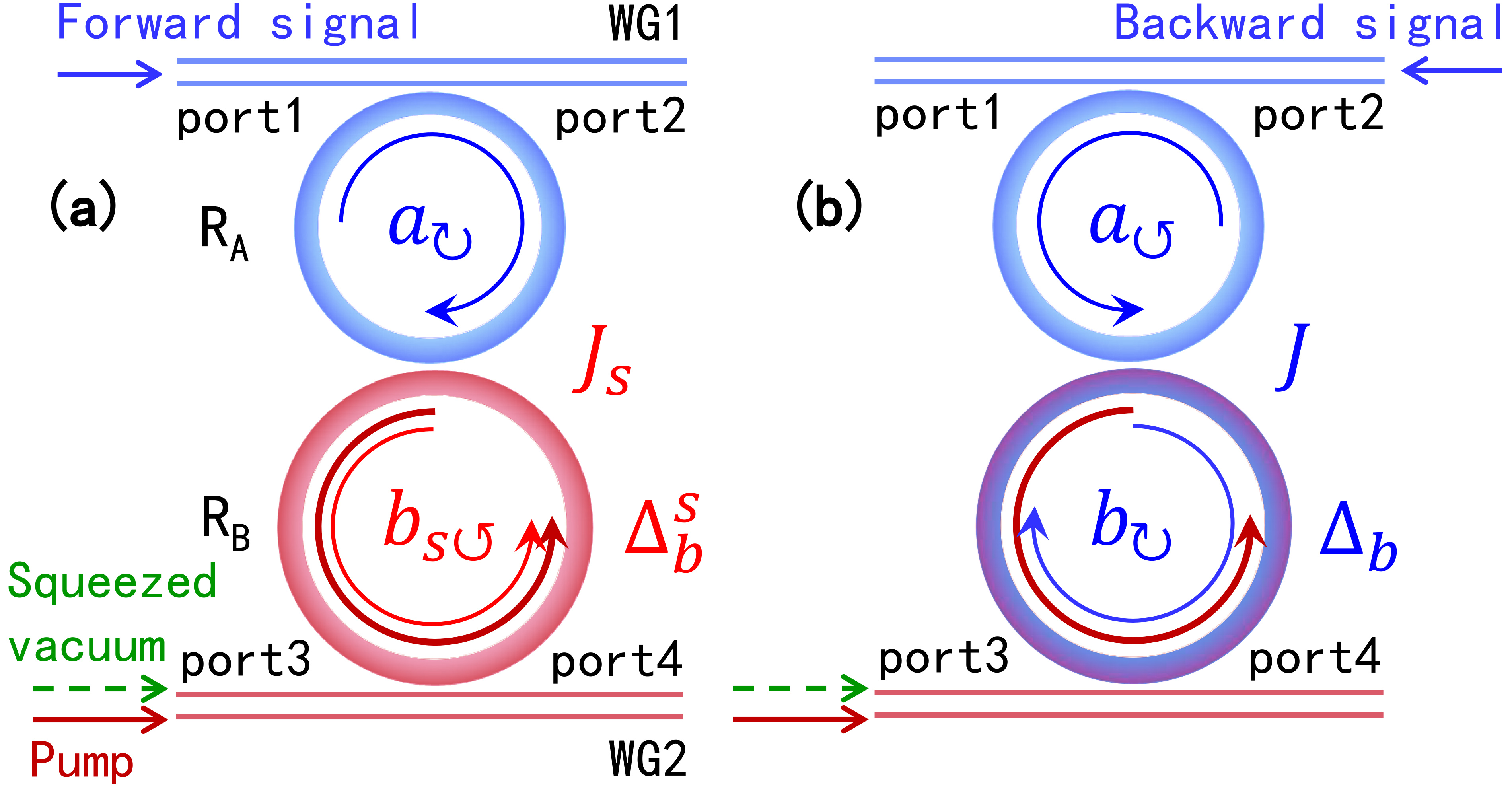} \\
  \caption{Schematic of an all-optical nonreciprocal system consisting of two microring resonators and two nearby optical waveguides.  To achieve \emph{classical light isolation}, a coherent pump field is applied to generate a CCW squeezing mode $b_{s_\circlearrowleft}$ in $\text{R}_\text{B}$. To achieve \emph{single-photon isolation}, a broadband squeezed-vacuum field is used to drive $\text{R}_\text{B}$.
  (a) A forward-input signal field excites a CW mode $a_{_\circlearrowright}$  in $\text{R}_\text{A}$, which interacts with the squeezed mode $b_{s_\circlearrowleft}$ with a coupling rate $J_s$.
  (b) A backward-input signal field stimulates a CCW mode $a_{_\circlearrowleft}$ in $\text{R}_\text{A}$. It couples to a CW bare mode $b_{_\circlearrowright}$ in $\text{R}_\text{B}$ with a rate $J$.}
  \label{fig:system}
 \end{figure}

 The schematic of our proposed system is depicted in Fig.~\ref{fig:system}. It consists of two coupled Whispering-Gallery mode microring resonators and two nearby optical waveguides. The resonators can be made of high-quality thin-film with $\chi^{(2)}$ nonlinearity, e.g., lithium niobate or aluminum nitride. Thus, the resonators supports the parametric nonlinear optical process.
 The resonator B ($\text{R}_\text{B}$) is pumped from port $3$ by a continuous wave coherent laser field with frequency $\omega_p$, amplitude $\alpha_p$, and phase $\theta_p$.
 This \emph{classical} pump generates a squeezing interaction with strength $\Omega_p$ for the  counterclockwise (CCW) mode $b_{_\circlearrowleft}$.
 Because of the directional phase-matching condition in the parametric nonlinear process, the mode $b_{_\circlearrowleft}$ is squeezed to a mode $b_{s_\circlearrowleft}$, but the clockwise (CW) mode $b_{_\circlearrowright}$ is unsqueezed.  The resonator A ($\text{R}_\text{A}$) slightly differs in size from $\text{R}_\text{B}$, such that the pump field cannot drive the parametric nonlinear process in the former. In this arrangement, we only need to consider mode squeezing in $\text{R}_\text{B}$. When a signal inputs to port 2, it ``sees'' a normal two-resonator system.

 Now we discuss how the pump laser modulates the inter-resonator interaction and causes the ONR.
 In the forward-input case, a signal field input to port $1$ excites the CW mode $a_{_\circlearrowright}$ in $\text{R}_\text{A}$. Because of the pump field, the mode $a_{_\circlearrowright}$ couples to the squeezed mode $b_{s_\circlearrowleft}$ with a rate $J_s$. The parametric nonlinear process related to $\Omega_p$ also causes a frequency shift to $b_{s_\circlearrowleft}$ with respect to the bare mode $b_{_\circlearrowleft}$.
 In comparison, in the backward-input case, a signal field from port $2$ excites the CCW mode $a_{_\circlearrowleft}$ in $\text{R}_\text{A}$. In this case, the pump field has no action on the CW mode $b_{_\circlearrowright}$. This case is equivalent to the unpumped system only consisting of two coupled resonators. Thus,  the mode $a_{_\circlearrowleft}$ interacts with the bare mode $b_{_\circlearrowright}$ with a unmodulated coupling rate $J$, rather than coupling to the squeezed mode.
Because the opposite input signal ``sees'' different system configurations, we can achieve ONR.

 In the frame rotating at frequency $\omega_p/2$, the Hamiltonian for the forward-input case reads
 \begin{equation} \label{eq:H_1}
  \begin{aligned}
       &\mathcal{H}_\text{fw} = \mathcal{H}_\text{A} + \mathcal{H}_\text{B} + \mathcal{H}_J\;,\\
       &\mathcal{H}_\text{A}/\hbar = \Delta_p^a  a_{_\circlearrowright}^\dagger a_{_\circlearrowright} + i\sqrt{2\kappa_\text{ex1}}(a_\text{in}a_{_\circlearrowright}^\dagger e^{-i\Delta_\text{in}t} - a_\text{in}^\dagger a_{_\circlearrowright} e^{i\Delta_\text{in}t})\;, \\
       &\mathcal{H}_\text{B}/\hbar = \Delta_p^b b_{_\circlearrowleft}^\dagger b_{_\circlearrowleft}  + \frac{\Omega_p}{2}(e^{-i\theta_p}b_{_\circlearrowleft}^{\dagger2}  + e^{i\theta_p}b_{_\circlearrowleft}^2 )\;,\\
       &\mathcal{H}_J/\hbar = J(a_{_\circlearrowright}^\dagger b_{_\circlearrowleft} + a_{_\circlearrowright} b_{_\circlearrowleft}^\dagger)\;,
  \end{aligned}
 \end{equation}
 where the detunings are $\Delta_p^{a/b}=\omega_{a/b}-\omega_p/2$ and $\Delta_\text{in}=\omega_\text{in}-\omega_p/2$, $\omega_{a/b}$ is the resonance frequency of $\text{R}_\text{A/B}$, 
 $\omega_\text{in}$ is the frequency of the signal mode $a_\text{in}$, and $\kappa_\text{ex1}$ is the external decay rate of $\text{R}_\text{A}$.
Experimentally, the pump field $\Omega_p$ is created by driving a mode $c_{_\circlearrowleft}$ of $\text{R}_\text{B}$ with an external field $\alpha_p$. To a good approximation, the dynamics of mode $c_{_\circlearrowleft}$ with resonance frequency $\omega_c$ can be modeled as a cavity driven by a laser with power $P_p$ and frequency $\omega_p \approx 2\omega_b$ via the Hamiltonian $H_\text{p} = \Delta_p^c c_{_\circlearrowleft}^\dagger c_{_\circlearrowleft} + i \sqrt{2\kappa_\text{ex2}^p} \alpha_p (c_{_\circlearrowleft}^\dagger -c_{_\circlearrowleft})$, where $\alpha_p = \sqrt{2\pi P_p/\hbar\omega_p}$, $\Delta_p^c = \omega_c -\omega_p$ and $\kappa_\text{ex2}^p$ is the external decay rate of the mode $c_{_\circlearrowleft}$. In steady state, we obtain $\langle c_{_\circlearrowleft} \rangle_\text{ss} =\sqrt{2\kappa_\text{ex2}^p} \alpha_p \big/ (i \Delta_p^c + \kappa_p)$, where $\kappa_p$ is the total decay rate of the mode $c_{_\circlearrowleft}$. In the mean-field approximation, we have $\Omega_p e^{-i\theta_p}= 2g \langle c_{_\circlearrowleft} \rangle_\text{ss}$, where $g$ is the single-photon nonlinear coupling strength of the parametric nonlinear process (see Supplemental Material~\cite{suppl.material}).

Applying the Bogoliubov squeezing transformation~\cite{QOBook.1997,QOBook.2012,prl.114.093602.2015,prl.120.093601.2018} $b_s=\cosh (r_p)b + e^{-i\theta_p}\sinh (r_p)b^\dagger$ with the squeezing parameter $r_p=\frac{1}{4}\ln(1+\beta\big/1-\beta)$ and $\beta=\Omega_p/\Delta_p^b$, we can transform the Hamiltonian $\mathcal{H}_\text{fw}$ to the squeezing picture.  We further apply the rotating-wave approximation $\Delta_p^a + \Delta_p^b\sqrt{1-\beta^2} \gg \sinh(r_p)J$ in the squeezing picture and neglect the counter-rotating terms. Then, the Hamiltonian in the frame rotating at frequency of $\Delta_\text{in}$ becomes
 \begin{equation}\label{eq:H1s}
  \begin{aligned}
   \mathcal{H}_\text{fw}^s/\hbar=&\Delta_a a_{_\circlearrowright}^\dagger a_{_\circlearrowright} + i\sqrt{2\kappa_\text{ex1}}(a_\text{in}a_{_\circlearrowright}^\dagger - a_\text{in}^\dagger a_{_\circlearrowright})\\
   &+\Delta_b^s b_{s_\circlearrowleft}^\dagger b_{s_\circlearrowleft}+J_s(a_{_\circlearrowright}^\dagger b_{s_\circlearrowleft} + b_{s_\circlearrowleft}^\dagger a_{_\circlearrowright})\;,
  \end{aligned}
 \end{equation}
 where $\Delta_a = \omega_a-\omega_\text{in}$, $\Delta_b^s=\Delta_p^{bs}-\Delta_\text{in}$, $\Delta_p^{bs}=\Delta_p^b\sqrt{1-\beta^2}$ and $J_s = \cosh (r_p)J$. The effective squeezed mode detuning $\Delta_b^s$ and the effective coupling rate $J_s$ are controlled by the pump field $\Omega_p$ and the corresponding detuning $\Delta_p^b$. When the ratio $\beta$ approaches unity, the rate $J_s$ between modes $a_{_\circlearrowright}$ and $b_{s_\circlearrowleft}$ is enhanced exponentially with respect to the bare coupling rate $J$ (see Supplemental Material~\cite{suppl.material} for details).
 Note that the pump field $\Omega_p$ also adds thermalization noise to mode $b_{s_\circlearrowleft}$.

 In the squeezing picture, the master equation of the system without a squeezed-vacuum driving takes the form
 \begin{equation}\label{eq:MEq1}
   \dot{\rho}_\text{fw} = -i[\mathcal{H}_\text{fw}^s,\rho_\text{fw}] + \mathcal{L}[L_a]\rho_\text{fw} + \mathcal{L}[L_{bs}]\rho_\text{fw} +  \mathcal{\pounds}_\text{n}[L_{bs}]\rho_\text{fw} \;,
 \end{equation}
 where $\rho_\text{fw}$ is the system density matrix, the term $\mathcal{L}[L_a] \rho_\text{fw}$ ($\mathcal{L}[L_{bs}]\rho_\text{fw}$) with operator $L_a = \sqrt{\kappa_a}a_{_\circlearrowright}$  ($L_{bs} = \sqrt{\kappa_b}b_{s_\circlearrowleft}$) describes the decay of the mode $a_{_\circlearrowright}$ ($b_{s_\circlearrowleft}$) with a rate $\kappa_a$ ($\kappa_b$), and $\mathcal{L}[o]\rho = 2o\rho o^\dagger - o^\dagger o \rho -\rho o^\dagger o$.
 %
 The term $\mathcal{\pounds}_\text{n}[L_{bs}]\rho_\text{fw}$ describes the effective thermalization noise of the mode $b_{s_\circlearrowleft}$ resulting from the \emph{classical coherent} pump. It is given by $\mathcal{\pounds}_\text{n}[L_{bs}]\rho_\text{fw} = N_p \mathcal{L}[L_{bs}]\rho_\text{fw}+ N_p \mathcal{L}[L_{bs}^\dagger]\rho_\text{fw} - M_p \mathcal{L}^\prime [L_{bs}]\rho_\text{fw} - M_p^* \mathcal{L}^\prime [L_{bs}^\dagger]\rho_\text{fw}$, where $N_p=\sinh^2(r_p)$, $M_p=e^{i\theta_p}\cosh(r_p)\sinh(r_p)$, and $\mathcal{L}^\prime[o]\rho = 2o\rho o - o o \rho -\rho o o$. This noise can limit the application of the system in the quantum regime.

To eliminate this squeezing-induced noise for achieving single-photon isolation, we can apply a broadband squeezed-vacuum field $\xi$ with squeezing parameter $r_e$ and reference phase $\theta_e$ to drive $\text{R}_\text{B}$ from port 3. When $r_e=r_p$ and $\theta_e + \theta_p =\pm n\pi~(n=1,3,5,...)$, the squeezing-induced noise associated with the term $\mathcal{\pounds}_\text{n}[L_{bs}]\rho_\text{fw}$ is cancelled~\cite{prl.114.093602.2015,prl.120.093601.2018,pra.100.062501.2019,suppl.material}.
In doing so, the mode $a_{_\circlearrowright}$ can couple to the squeezed mode $b_{s_\circlearrowleft}$ coherently without additional noise, just as a simple linear resonator system.
A squeezed-vacuum field with $\tera\hertz$ bandwidth has been realized via optical parametric amplification~\cite{APL.Photonics.036104.2020}. Thus, the bandwidth of the squeezed vacuum is larger than the linewidth of the resonators in our system. In this case, the squeezed mode is equivalently coupled to a normal vacuum bath.
The decay rate of the squeezed mode equals to that of the bare mode such that $\kappa_{bs} = \kappa_b$.

For the backward-input case, the Hamiltonian reads
 \begin{equation}\label{eq:H2}
  \begin{aligned}
   \mathcal{H}_\text{bw}/\hbar=&\Delta_a a_{_\circlearrowleft}^\dagger a_{_\circlearrowleft} + i\sqrt{2\kappa_\text{ex1}}(a_\text{in}a_{_\circlearrowleft}^\dagger - a_\text{in}^\dagger a_{_\circlearrowleft})\ \\
   &+\Delta_b b_{_\circlearrowright}^\dagger b_{_\circlearrowright}+J(a_{_\circlearrowleft}^\dagger b_{_\circlearrowright} + b_{_\circlearrowright}^\dagger a_{_\circlearrowleft}) \;,
  \end{aligned}
 \end{equation}
 where $\Delta_b = \omega_b-\omega_\text{in}$. Comparing Eq.~\eqref{eq:H2} with Eq.~\eqref{eq:H1s}, we can see that the intermode detuning and coupling in these two Hamiltonians can be very different due to the directional quantum squeezing.
The dynamics of the system is governed by the master equation of two coupling resonators
 \begin{equation} \label{eq:MEq3}
  \dot{\rho}_\text{bw} = -i[\mathcal{H}_\text{bw},\rho_\text{bw}] + \mathcal{L}[L_a]\rho_\text{bw} + \mathcal{L}[L_b]\rho_\text{bw} \;,
 \end{equation}
 where $\rho_\text{bw}$ is the density matrix of the system, $L_a = \sqrt{\kappa_a}a_{_\circlearrowleft}$, and $L_b = \sqrt{\kappa_b}b_{_\circlearrowright}$. Note that in this case the squeezed-vacuum field has no influence on the dynamics.
 Thus, we can attain strong ONR by parametrically pumping the mode $b_{_\circlearrowleft}$.

 According to the input-output relation~\cite{pra.31.1985}, for an input field $a_\text{in}$, we have $a_\text{out}=a_{\text{in}} - \sqrt{2\kappa_{\text{ex1}}} a$ and $b_\text{out} = \sqrt{2\kappa_\text{ex2}} b$, where $\kappa_\text{ex2}$ is the external decay rate of $\text{R}_\text{B}$.
 The transmissions are defined as $T_{12/21} = \langle a_\text{out}^\dagger a_\text{out}\rangle \big/ \langle a_\text{in}^\dagger a_\text{in} \rangle$ and $T_{23} = \langle b_\text{out}^\dagger b_\text{out} \rangle \big/ \langle a_\text{in}^\dagger a_\text{in}\rangle$, where $T_{ij}$ is the transmission from port $i$ to port $j$, with $i,~j = 1,~2,~3$.
 Numerically solving Eqs.~(\ref{eq:MEq1},\ref{eq:MEq3}) and using $\alpha_\text{in} \equiv \langle a_\text{in} \rangle =  \sqrt{2\pi P_\text{in}/\hbar\omega_\text{in}}$ with the signal power $P_\text{in}$, we can obtain the steady-state solutions for $\langle a_\text{out}^\dagger a_\text{out} \rangle_\text{ss}$ and $\langle b_\text{out} ^\dagger b_\text{out}\rangle_\text{ss}$, and the steady-state transmissions.

 We can also analytically derive the transmission from the Langevin equations of motion. To consider the squeezing-induced noise, we truncate the Langevin equation to second-order nonlinear terms of operators and obtain
 %
 \begin{equation} 
  \begin{aligned}
   &\frac{\mathrm{d}}{\mathrm{d}t} a_x = -\left(i\Delta_a + \kappa_a\right)a_x + \sqrt{2\kappa_\text{ex1}}a_\text{in} - i J_x b_x\;, \\
   &\frac{\mathrm{d}}{\mathrm{d}t} b_x = -\left(i\Delta_b^x+\kappa_b\right)b_x - iJ_x a_x \;, \\
   &\frac{\mathrm{d}}{\mathrm{d}t}  a_x^\dagger b_x   =
   \left(i\Delta_{ab}^x-\kappa_{ab} \right)a_x^\dagger b_x + \sqrt{2\kappa_\text{ex1}}a_\text{in}^\dagger b_x - i J_x \Xi\;,\\
   &\frac{\mathrm{d}}{\mathrm{d}t}  b_x^\dagger b_x   =
   i J_x \left(a_x^\dagger b_x - a_x b_x^\dagger \right) - 2\kappa_b b_x^\dagger b_x + \Psi_\text{noise} \;,\\
   &\frac{\mathrm{d}}{\mathrm{d}t}  a_x^{\dagger} a_x   =
  - i J_x \left(a_x^\dagger b_x - a_x b_x^\dagger \right) + \sqrt{2\kappa_\text{ex1}}\left(a_\text{in}a_x^\dagger + a_\text{in}^\dagger a_x \right)\\
   &~~~~~~~~~~~~~~~  - 2\kappa_a a_x^\dagger a_x \;,
  \end{aligned}
 \end{equation}
 where the squeezing-induced noise $\Psi_\text{noise}= 2\sinh^2(r_p)\kappa_b$ is present in the forward-input case and plays the role of a thermal bath. In the backward-input case, $\Psi_\text{noise}$ is absent.
 Here, we have used $\Delta_{ab}^x = \Delta_a - \Delta_b^x$, $\kappa_{ab} = \kappa_a + \kappa_b$ and $\Xi = a_x^\dagger a_x - b_x^\dagger b_x$.
 By solving the Langevin equation and using the input-output relations, we obtain the steady-state transmissions:
 \begin{subequations}\label{eq:T}
  \begin{align}
    T_{12} &=  \frac{J_s^4+2 \zeta_x J_s^2 + \Lambda_x}{\mathcal{G}_x} + \frac{2\kappa_\text{ex1}\mathcal{N}_\text{noise}}{|\alpha_\text{in}|^2}\;, \label{eq:T1a} \\
    T_{21} &=  \frac{J^4+2 \zeta_x J^2 + \Lambda_x}{\mathcal{G}_x}\;, \label{eq:T1b}\\
    T_{23} &= \frac{4\kappa_\text{ex1}\kappa_\text{ex2}J^2}{\mathcal{G}_x} \;,\label{eq:T23}
  \end{align}
 \end{subequations}
 where $\mathcal{N}_\text{noise} = \kappa_b(\kappa_a+\kappa_b)\sinh^2(r_p)J_s^2/\mathcal{Q}_s$ is the number of noise-related photons, $\mathcal{Q}_s=J_s^2(\kappa_a+\kappa_b)^2+\kappa_a\kappa_b[(\kappa_a+\kappa_b)^2+{\Delta_{ab}^s}^2]$, $\mathcal{G}_x=J_x^4+2J_x^2(\kappa_a \kappa_b - \Delta_a \Delta_b^x)+(\kappa_a^2+\Delta_a^2)(\kappa_b^2+{\Delta_b^x}^2)$, $\zeta_x=\kappa_a \kappa_b - 2\kappa_b \kappa_\text{ex1}-\Delta_a \Delta_b^x$, and $\Lambda_x=[(\kappa_a-2\kappa_\text{ex1})^2+\Delta_a^2](\kappa_b^2+{\Delta_b^x}^2)$.
In the calculations, we need to replace $a_x$ with $a_{_\circlearrowright}$, $b_x$ with $b_{s_\circlearrowleft}$, $J_x$ with $J_s$, and $\Delta_{b}^x$ with $\Delta_{b}^s$ ($a_x$ with $a_{_\circlearrowleft}$, $b_x$ with $b_{_\circlearrowright}$, $J_x$ with $J$, and $\Delta_{b}^x$ with $\Delta_{b}$) for the forward-input (backward-input) case (see Supplemental Material~\cite{suppl.material}).

By applying the squeezed-vacuum field to cancel the noise $\Psi_\text{noise}$, yielding $\Psi_\text{noise}= 0$ and $\mathcal{N}_\text{noise}=0$. The steady-state transmissions become
 \begin{equation} \label{eq:Tsv}
 T_{12}^\text{sv} = (J_s^4+2 \zeta_s J_s^2+\Lambda_s) \big/ \mathcal{G}_s \;, \quad T_{21}^\text{sv} = T_{21} \;, \quad
 T_{23}^\text{sv}=T_{23} \;.
 \end{equation}
The noise-free transmission $T_{12}^\text{sv}$ is the limitation of $T_{12}$ for a classical large input $\alpha_\text{in}$ and also valid for single-photon pulses. Thus, we can achieve ONR in both the classical and quantum regimes.

Below, we will show two different mechanisms to achieve a strong ONR dependence on the bare mode coupling rate $J$. When $J<\kappa_a, \kappa_b$, the two resonators originally have no NMS. We use the pump field to induce the NMS between $a_{_\circlearrowright}$ and $b_{s_\circlearrowleft}$, namely the NMS scenario. For $J \gg \kappa_a, \kappa_b$  resulting in the NMS between the bare modes, we use the pump field to significantly shift the resonance frequency of the mode $b_{s_\circlearrowleft}$. We call this mechanism the mode resonance shift (MRS) scenario.

 \begin{figure}[ht]
  \centering
  \includegraphics[width=0.99\linewidth]{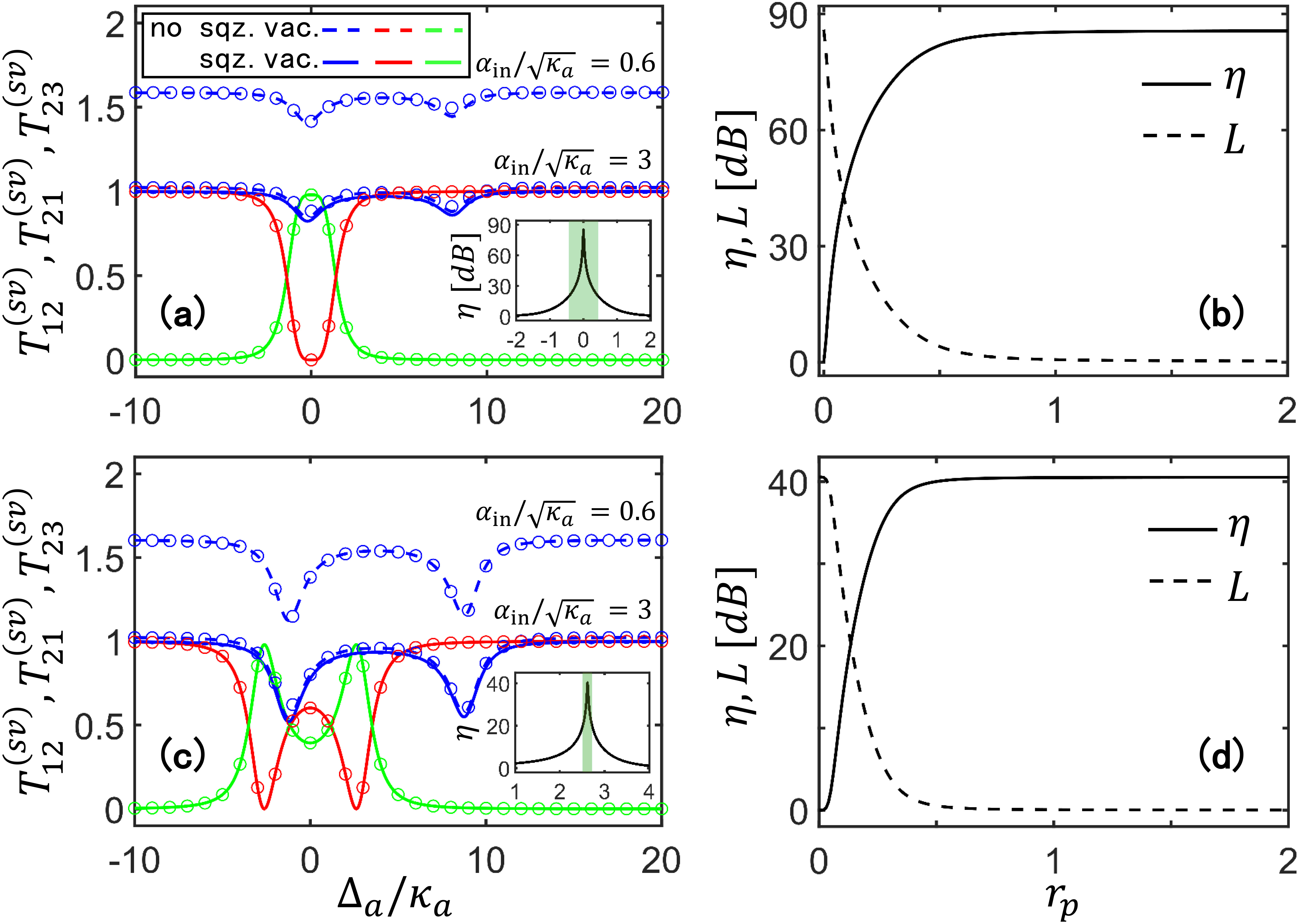} \\
  \caption{ (a) and (c) Steady-state transmission and isolation (inset) versus $\Delta_a$.  (b) and (d) Isolation and insertion loss versus $r_p$. (a) and (b) for the NMS scenario ($J/\kappa_a=0.99$), (c) and (d) for the MRS scenario ($J/\kappa_a=2.8$). Curves and circles are for analytical and numerical transmissions of $T_{12}^{(\text{sv})}$ (blue), $T_{21}^{(\text{sv})}$ (red) and $T_{23}^{(\text{sv})}$ (green), respectively. Solid (dashed) curves are for analytical transmissions $T_{ij}^\text{sv}$ ($T_{ij}$) with (without) the squeezed-vacuum field. Circles show the corresponding numerical transmissions $T_{ij}$.
  Other parameters are: $\Delta_a=\Delta_b$, $\kappa_a=\kappa_b$, $\kappa_\text{ex1,2}/\kappa_a=0.99$; in (a) $\Delta_p^b/\kappa_a=10.3$, $\Omega_p/\kappa_a=10$, yielding $r_p\sim1.05$; in (c) $\Delta_p^b/\kappa_a=15$, $\Omega_p/\kappa_a=13$, yielding $r_p\sim0.66$; in (b) $\Delta_a = 0$ and $\Delta_p^b = 10\sinh(r_p)$; in (d) $\Delta_a/\kappa_a = 2.62$ and $\Delta_p^b = 30\sinh(r_p)$.}
  \label{fig:figure2}
 \end{figure}

Our system can realize an optical diode with transmission $T_{12} \gg T_{21}$ even with only the \emph{classical coherent} pump $\Omega_p$. For a weak input signal, e.g. $\alpha_\text{in}/\sqrt{\kappa_a} =0.6$, the transmission $T_{12}$ (blue dashed curves) can be larger than unity, because the pump-induced noise can thermalize the squeezed mode. As the intensity of the incident signal increases to $\alpha_\text{in}/\sqrt{\kappa_a} > 3$, corresponding to a weak signal including $9$ photons within the resonator decay time, the input signal overwhelms the pump-induced noise. In this case, the transmission $T_{12}$ in the classical regime approximates the noise-free transmission $T_{12}^\text{sv}$ (blue solid curves) in the quantum regime. The analytical transmissions are in excellent agreement with the numerical results, see Figs.~\ref{fig:figure2}(a,c). Below, we focus on the classical case with a large signal input or the noise-free transmission when the squeezed-vacuum field is present.
 In the NMS scenario, we choose $J=0.99\kappa_a$ for the critical coupling condition, such that the transmission $T_{21}$ vanishes in the backward-input case. In comparison, in the forward-input case, the squeezing interaction enhances the coupling strength $J_s$ to much larger than the system decay rates and leads to a large NMS, see Fig.~\ref{fig:figure2}(a). Therefore, we obtain $T_{12}^\text{sv} \approx 83.1\%$ and $T_{21}^\text{sv} \approx 0$ at $\Delta_a = 0$, corresponding to an isolation ratio $\eta = 10\log(T_{12}^\text{sv}/T_{21}^\text{sv}) \approx 85.1~\deci\bel$, insertion loss ${L}=-10\log(T_{12}^\text{sv}) \approx 0.80~\deci\bel$. The bandwidth for an isolation ratio larger than $20~\deci\bel$ is about $0.86\kappa_a$, see the inset of Fig.~\ref{fig:figure2}(a).
 In the MRS scenario, we take $J=2.8\kappa_a$ for example. The transmission spectrum splits in the backward-input case. In the forward-input case, the resonance frequency of the squeezed mode is shifted and thus detuned from the mode $a_{_\circlearrowright}$. The transmission spectrum is shifted with respect to the backward-input transmission, see Fig.~\ref{fig:figure2}(c). As a result, we attain $T_{12}^\text{sv} \approx 92.8\%$ and $T_{21}^\text{sv} \approx 0$ at $\Delta_a/\kappa_a = 2.62$, corresponding to $\eta \approx 40.3~\deci\bel$, ${L} \approx 0.32~\deci\bel$. The bandwidth for $\eta \geqslant 20~\deci\bel$ is about $0.21\kappa_a$, see the inset of Fig.~\ref{fig:figure2}(c).
 The isolation ratio and the insertion loss improve with $r_p$ and reach stable values when $r_p > 0.6$, see Figs.~\ref{fig:figure2}(b,d).

Similar to a commercial circulator, our proposed system can also function as a quasi-circulator with two inputs and three outputs, allowing photon flow along the direction $1 \rightarrow 2 \rightarrow 3$~\cite{adr.2.2021}, see Figs.~\ref{fig:figure2}(a,c). Here, we focus on  the fidelity and the insertion loss. To evaluate the quasi-circulator performance, we calculate the average fidelity as $\mathcal{F} = \text{Tr} \left[ \tilde{T} {T^\text{id}}^T \right] \bigg/ \text{Tr} \left[ T^\text{id} {T^\text{id}}^T \right]$~\cite{science.354.2016,prl.121.203602.2018}, where $T^\text{id}$ is the transmission matrix for an ideal three-port quasi-circulator~\cite{adr.2.2021}, and $\tilde{T} = T_{ij}/\Upsilon_i$, with $\Upsilon_i = \textstyle \sum_{j}T_{ij}$. We define the average insertion loss as $\tilde{{L}} = -10 \log[(T_{12} + T_{23})/2]$. The transmissions $T_{12}$ and $T_{21}$ are the same as the diode. For an input $\alpha_\text{in}/\sqrt{\kappa_a} >3$, the pump-induced noise is negligible.
 In the NMS scenario, we achieve $T_{23} \approx 98.0\%$ at $\Delta_a = 0$, corresponding to $\tilde{{L}} \approx 0.43~\deci\bel$. We also obtain similar performance in the MRS scenario with $T_{23} \approx 98.0\%$ at $\Delta_a/\kappa_a = 2.62$, corresponding to $\tilde{{L}}\approx 0.20~\deci\bel$. In both scenarios, we have $\mathcal{F} \approx 1$.

 Our system can work as a single-photon quasi-circulator when the squeezing-vacuum field is applied to cancel the noise term. We evaluate its performance for a single-photon wavepacket input to ports $1$ and $2$ simultaneously by solving a quantum cascaded system~\cite{Gardiner1993,Carmichael1993,njp.14.2012} (also Supplemental Material~\cite{suppl.material}). We consider a Gaussian-like single-photon pulse with duration $2\pi \times 6\kappa_a^{-1}$. In comparison with the case without the squeezed-vacuum field, the performance of our quasi-circulator only slightly decreases. The fidelity is still very high, $\mathcal{F}\geq 98.7\%$. The insertion loss remains unchanged due to the large bandwidth in the NMS scenario. In comparison, it reduces to $\tilde{{L}}\approx 0.26~\deci\bel$ at $\Delta_a / \kappa_a= 2.62$ in the MRS scenario.
Therefore, our diode and quasi-circulator can work in both the classical and quantum regimes.

 When the pump laser with power $P_p$ is present, corresponding to ``ON'' (absent, corresponding to ``OFF''), our device can switch on and off the transmission of a stronger signal laser from $T_{12}^{\text{on}}$ to $T_{12}^{\text{off}}$ according to Eq.~\ref{eq:T}(a), implying an all-optical nonreciprocal transistor.
 As the crucial role of electronic transistor in electric computers, optical transistors are essential for optical information processing~\cite{science.361.2018,science.341.768.2013,nature.460.76.2006,nat.commun.4.1778.2013}.
 We define the gain of the transistor as $G = \frac{P_\text{in}}{P_p}\Delta T $, with $\Delta T= T_{12}^\text{on} - T_{12}^\text{off}$~\cite{science.361.2018}. The gain increases linearly with the signal power.
 When we fix the pump strength, e.g. $\Omega_p/\kappa_a = 10$, in the NMS scenario, the gain of the transistor can reach $G>1$ ($G>100$) when $|\alpha_\text{in}|^2/\kappa_a > 6.1 \times10^7$ ($|\alpha_\text{in}|^2/\kappa_a>6.1 \times10^9$). In the MRS scenario, we can obtain the same gain $G>1$ ($G>100$) by taking $\Omega_p/\kappa_a = 13$ and applying a slightly larger pump power $|\alpha_\text{in}|^2/\kappa_a > 9.2 \times 10^7 $ ($|\alpha_\text{in}|^2/\kappa_a>9.2 \times10^9$).
 Unlike the forward-input case, the transmission $T_{21}$ in the backward-input case is independent of the pump power and always vanishing small. Clearly, our optical transistor is nonreciprocal.
 \begin{figure}[ht]
  \centering
  \includegraphics[width=0.99\linewidth]{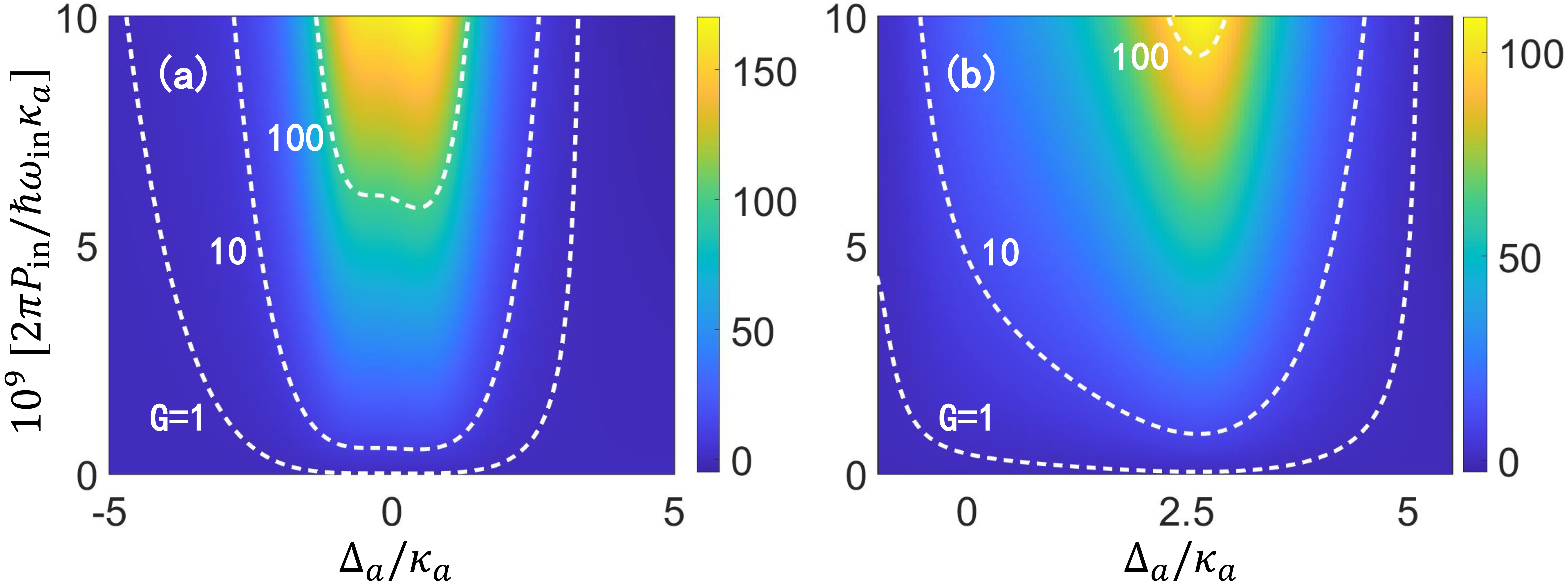} \\
  \caption{Gain of the transistor versus detuning $\Delta_a$ and $P_\text{in}$. (a) The NMS scenario with $J/\kappa_a=0.99$, $\Delta_p^b/\kappa_a=10.3$, $\Omega_p/\kappa_a=10$. (b) The MRS scenario with $J/\kappa_a=2.8$, $\Delta_p^b/\kappa_a=15$, $\Omega_p/\kappa_a=13$. Other parameters are: $\Delta_a=\Delta_b$, $\kappa_a=\kappa_b$, $\kappa_\text{ex1,2}/\kappa_a=0.99$, $g =10^{-3}\kappa_a$.}
  \label{fig:figure3}
 \end{figure}

 Lithium-niobate-based microring resonators provide an excellent platform for our proposal, thanks to their large $\chi^{(2)}$~\cite{oe.25.2017,njp.22.2020,prl.125.263602.2020} and high optical quality factors up to $Q \sim 10^7$~\cite{optica.4.2017,optica.6.2019}.
 Assuming an experimentally available intrinsic quality factor $Q_i = 8 \times 10^6$, the resonance frequency of the signal field $\omega_{a/b}/2\pi  = 193.4~\tera\hertz$ and the pump field frequency  $\omega_p = 2\omega_a$  for the resonators, we obtain the intrinsic loss rate $\kappa_i/2\pi =24.2~\mega\hertz$. By choosing an experimentally available gap~\cite{prapplied.15.2021}, we can select feasible values for the external decay rates and the coupling rate: $\kappa_\text{ex1}= \kappa_\text{ex2} \approx 2\pi \times 2.40~\giga\hertz$, $\kappa_{a} = \kappa_{b} \approx 2\pi \times 2.42~\giga\hertz$ and $J = 0.99 \kappa_a \approx 2\pi \times 2.40~\giga\hertz$ in the NMS scenario (or $J = 2.8\kappa_a  \approx 2\pi \times 6.78~\giga\hertz$ in the MRS scenario). Hence, the nonreciprocal bandwidth for $\eta \geq 20~\deci\bel$ can reach $0.86\kappa_a/2\pi \approx  2.08 ~\giga\hertz$ around $\Delta_a =0$ (or $0.21\kappa_a/2\pi \approx 0.51 ~\giga\hertz$ around $\Delta_a =2.62 \kappa_a$).
 The power of a pump field is given by $P_p = {\hbar \omega_p \kappa_p^2 \Omega_p^2} / {16\pi g^2 \kappa_{\text{ex2}}^p}$~\cite{suppl.material}. The rate of $g/2\pi= 2.35~\mega\hertz$ is available for Lithium-niobate-based microresonators~\cite{optica.7.1654.2020}.
 Thus, the pump power $P_p \approx 16.6~\milli\watt$ (or $28.0~\milli\watt$) yields to $\Omega_p/\kappa_a=10$ (or $13$). At the detuning $\Delta_a/\kappa_a = 0$ (or $2.62$), we can induce an optical transistor with $G>1$ for a signal power $P_\text{in} \approx 18.9~\milli\watt$ in the NMS scenario (or $28.5~\milli\watt$ in the MRS scenario).

 We have proposed a chip-compatible magnetic-free system to realize an optical diode, a quasi-circulator and a nonreciprocal optical transistor. The optical diode and circulator can work \emph{in the classical regime under a coherent pump and also in the quantum regime when a squeezed-vacuum field is applied}.
 Our method only needs two-mode phase matching in one resonator. It greatly simplifies its experimental implementation, in comparison with a non-degenerate three-mode matching counterpart~\cite{prl.117.2016, prl.126.133601.2021}. In particular, our work proposed the nonreciprocal optical transistor switching a strong signal with a weak control field. Such unconventional transistor cannot be realized in the configuration of Ref.~\cite{prl.117.2016} because a strong signal will cause NMS for the pump field.

\section*{Acknowledgements}
This work was supported by the National Key R\&D Program of China (Grants No. 2017YFA0303703, No. 2019YFA0308700), the National Natural Science Foundation of China (Grant Nos. 11874212, 11890704, 11690031), the Fundamental Research Funds for the Central Universities (Grant No. 021314380095), the Program for Innovative Talents and Entrepreneurs in Jiangsu, and the Excellent Research Program of Nanjing University (Grant No. ZYJH002).
F.N. is supported by the Nippon Telegraph and Telephone Corporation (NTT) Research, the Japan Science and Technology Agency (JST) [via the Quantum Leap Flagship Program (Q-LEAP), the Moonshot R\&D Grant Number JPMJMS2061, and the Centers of Research Excellence in Science and Technology (CREST) Grant No. JPMJCR1676], the Japan Society for the Promotion of Science (JSPS) [via the Grants-in-Aid for Scientific Research (KAKENHI) Grant No. JP20H00134 and the JSPS-RFBR Grant No. JPJSBP120194828], the Army Research Office (ARO) (Grant No. W911NF-18-1-0358), the Asian Office of Aerospace Research and Development (AOARD) (via Grant No. FA2386-20-1-4069), and the Foundational Questions Institute Fund (FQXi) via Grant No. FQXi-IAF19-06. We thank the High Performance Computing Center of Nanjing University for doing the numerical calculations on its blade cluster system.

%

\end{document}


\preprint{\hfill\parbox[b]{0.3\hsize}{ }}

\title{
{\it Supplemental Material for}\\
``Quantum Squeezing Induced Optical Nonreciprocity''}

\author{Lei Tang,$^{1}${
\begin{tabular}{ll}
\end{tabular}}
Jiangshan Tang,$^{1}$
Mingyuan Chen,$^{1}$
Franco Nori,$^{2,3}$
Min Xiao,$^{1,4}$
and Keyu Xia$^{1,5}$$^*$}
\affiliation{\it
$^1$College of Engineering and Applied Sciences, National Laboratory of Solid State Microstructures, and Collaborative Innovation Center of Advanced Microstructures, Nanjing University, Nanjing 210093, China\\
$^2$RIKEN Quantum Computing Center, RIKEN Cluster for Pioneering Research, Wako-shi, Saitama 351-0198, Japan\\
$^3$Physics Department, The University of Michigan, Ann Arbor, Michigan 48109-1040, USA\\
$^4$Department of Physics, University of Arkansas, Fayetteville, Arkansas 72701, USA \\
$^5$Jiangsu Key Laboratory of Artificial Functional Materials, Nanjing University, Nanjing 210023, China
}
\date{\today}

\maketitle
 In this Supplemental Material, we first present more details about the Bogoliubov squeezing transformation for the forward-input Hamiltonian.
 Then, we derive the master equation of the system including the squeezing-induced noise for calculating the steady-state transmission of the classical light. We also show more details of the elimination of the squeezing-induced noise and derive the master equation of the quantum cascaded system without the noise, to attain the transmission matrix of single-photon pulses.
 After that, we present the quantum Langevin equations of the system, in the cases with and without the squeezing-induced noise, for deriving the steady-state analytical transmissions.
 Finally, we derive the relation of the pump strength $\Omega_p$ and the pump power $P_p$.
 \newline


\section{Squeezing parameter in the forward-input case}
 For convenience, we recall the Hamiltonian given by Eq.~(1) in the main text for the forward-input case
 \begin{equation} \label{eq:H1}
  \begin{aligned}
   &\mathcal{H}_\text{fw} = \mathcal{H}_\text{A} + \mathcal{H}_\text{B} + \mathcal{H}_J\;,  \\
   &\mathcal{H}_\text{A} /\hbar = \Delta_p^a  a_{_\circlearrowright}^\dagger a_{_\circlearrowright} + i\sqrt{2\kappa_\text{ex1}}(a_\text{in}a_{_\circlearrowright}^\dagger e^{-i\Delta_\text{in}t} - a_\text{in}^\dagger a_{_\circlearrowright} e^{i\Delta_\text{in}t})\;,  \\
   &\mathcal{H}_\text{B} /\hbar = \Delta_p^b b_{_\circlearrowleft}^\dagger b_{_\circlearrowleft}  + \frac{\Omega_p}{2}(e^{-i\theta_p}b_{_\circlearrowleft}^{\dagger2}  + e^{i\theta_p}b_{_\circlearrowleft}^2 )\;,  \\
   &\mathcal{H}_J /\hbar = J(a_{_\circlearrowright}^\dagger b_{_\circlearrowleft} + a_{_\circlearrowright} b_{_\circlearrowleft}^\dagger) \;,
  \end{aligned}
 \end{equation}
 where the detuning $\Delta_p^{a/b}=\omega_{a/b}-\omega_p/2$ and $\Delta_\text{in}=\omega_\text{in}-\omega_p/2$.
 By applying the Bogoliubov squeezing transformation $b_s=\cosh (r_p)b + e^{-i\theta_p}\sinh (r_p)b^\dagger$~\cite{QOBook.1997, prl.114.093602.2015, prl.120.093601.2018},  we obtain the Hamiltonian in the squeezing picture. Under the condition $\Delta_p^a + \Delta_p^b\sqrt{1-\beta^2} \gg \sinh(r_p)J$, with $\beta=\Omega_p/\Delta_p^b$, we can apply the rotating-wave approximation and neglect the counter-rotating terms. Then, the Hamiltonian in the squeezing picture is given by
 \begin{equation}\label{eq:H1s}
  \begin{aligned}
   &\mathcal{H}_\text{fw}^s = \mathcal{H}_\text{A}^s + \mathcal{H}_\text{B}^s + \mathcal{H}_J^s \;, \\
   &\mathcal{H}_\text{A}^s /\hbar = \Delta_p^a  a_{_\circlearrowright}^\dagger a_{_\circlearrowright} + i\sqrt{2\kappa_\text{ex1}}(a_\text{in}a_{_\circlearrowright}^\dagger e^{-i\Delta_\text{in}t} - a_\text{in}^\dagger a_{_\circlearrowright} e^{i\Delta_\text{in}t})\;,  \\
   &\mathcal{H}_{\text{B}}^s /\hbar = \Delta_p^{bs} b_{s_\circlearrowleft}^{\dagger} b_{s_\circlearrowleft}\;, \\
   &\mathcal{H}_J^s /\hbar = J_s(a_{_\circlearrowright}^\dagger b_{s_\circlearrowleft} + a_{_\circlearrowright}  b_{s_\circlearrowleft}^{\dagger}) \;.
  \end{aligned}
 \end{equation}
 Here, after the Bogoliubov squeezing transformation, the bare mode $b_{_\circlearrowleft}$ is transformed to the squeezed mode $b_{s_\circlearrowleft}$ with detuning
 \begin{equation}
  \Delta_p^{bs}=\Delta_p^b\sqrt{1-\beta^2} \;,
 \end{equation}
 and squeezing parameter
 \begin{equation}
   r_p=\frac{1}{4}\ln\frac{1+\beta}{1-\beta}\;.
 \end{equation}
 Consequently, for $r_p>0$, the coupling rate between the bare mode $a_{_\circlearrowright}$ and the squeezed mode $b_{s_\circlearrowleft}$ is exponentially enhanced~\cite{prl.114.093602.2015, prl.120.093601.2018}, and given by
  \begin{equation}
   J_s = \cosh(r_p)J\;.
 \end{equation}
 In a frame rotating at frequency $\Delta_\text{in}$, the Hamiltonian Eq.~\eqref{eq:H1s} can be rewritten as
 \begin{equation}\label{eq:H1ss}
   \mathcal{H}_\text{fw}^s /\hbar = \Delta_a a_{_\circlearrowright}^\dagger a_{_\circlearrowright} + i\sqrt{2\kappa_\text{ex1}}(a_\text{in}a_{_\circlearrowright}^\dagger - a_\text{in}^\dagger a_{_\circlearrowright}) + \Delta_b^s b_{s_\circlearrowleft}^{\dagger} b_{s_\circlearrowleft}+J_s(a_{_\circlearrowright}^\dagger b_{s_\circlearrowleft} + b_{s_\circlearrowleft}^{\dagger}a_{_\circlearrowright}) \;,
 \end{equation}
 where $\Delta_a = \Delta_p^a - \Delta_\text{in} = \omega_a-\omega_\text{in}$, $\Delta_b^s=\Delta_p^{bs}-\Delta_\text{in}$.

 \begin{figure}[ht]
  \centering
  \includegraphics[width=0.8\linewidth]{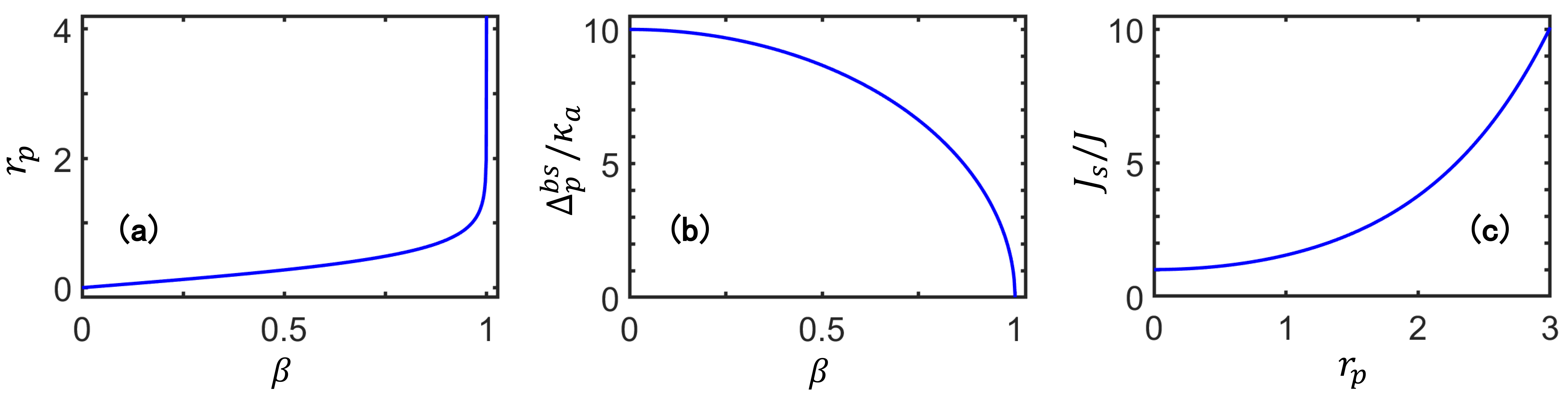} \\
  \caption{ (a) The squeezing parameter $r_p$ versus $\beta \in [0,1)$. (b) The detuning $\Delta_p^{bs}$ versus $\beta \in [0,1)$, $\Delta_p^b/\kappa_a=10$. (c) The enhanced coupling rate $J_s$ varying with the squeezing parameter $r_p\in[0,3]$. }
  \label{fig:squeezing_paramter}
 \end{figure}

 As depicted in Fig.~S\ref{fig:squeezing_paramter}, when the ratio $\beta$ approaches $1$, the squeezing parameter increases greatly and the detuning $\Delta_p^{bs}$ decreases to $0$. Furthermore, increasing the squeezing parameter $r_p$ leads to an exponential enhancement~\cite{prl.114.093602.2015, prl.120.093601.2018} of the coupling rate $J_s$.

 \section{Master equation}
 \subsection{Squeezing-induced noise and classical-light input}
 For our system in the forward-input case, the dynamics of the system coupling to a normal vacuum reservoir is described by a standard master equation
 \begin{equation} \label{eq:MEq1}
  \frac{\mathrm{d}}{\mathrm{d}t}{\rho_\text{fw}}=-i[\mathcal{H}_\text{fw},\rho_\text{fw}] + \mathcal{L}[L_a]\rho_\text{fw} + \mathcal{L}[L_b]\rho_\text{fw} \;,
 \end{equation}
 where $\rho_\text{fw}$ is the density matrix of the system in the forward-input case, $\mathcal{H}_\text{fw}$ is the Hamiltonian given by Eq.~\eqref{eq:H1}, and the operators $L_a = \sqrt{\kappa_a}a_{_\circlearrowright}$ and $L_b = \sqrt{\kappa_b}b_{_\circlearrowleft}$ describe the decay of the resonator A ($\text{R}_\text{A}$) and B ($\text{R}_\text{B}$) with rates $\kappa_a$ and $\kappa_b$, respectively. Moreover, $\mathcal{L}[o]\rho = 2o\rho o^\dagger - o^\dagger o \rho -\rho o^\dagger o$.

 Next, we apply the Bogoliubov squeezing transformation, $b=\cosh(r_p)b_s-e^{-i\theta_p}\sinh(r_p)b_s^\dagger$, to respectively replace the Hamiltonian $\mathcal{H}_\text{fw}$ with $\mathcal{H}_\text{fw}^s$ [Eq.~\eqref{eq:H1ss}] and $L_b$ with $L_b^s = \sqrt{\kappa_b}[\cosh(r_p)b_{s_\circlearrowleft} - e^{-i\theta_p}\sinh(r_p)b_{s_\circlearrowleft}^\dagger]$. Therefore, we can rewrite Eq.~\eqref{eq:MEq1} in the squeezed picture as
 \begin{equation}\label{eq:MEq1s}
  \begin{aligned}
   \frac{\mathrm{d}}{\mathrm{d}t} \rho_\text{fw} &=-i[\mathcal{H}_\text{fw}^s,\rho_\text{fw}] + \mathcal{L}[L_a]\rho_\text{fw} + \mathcal{L}[L_b^s]\rho_\text{fw} \\
   &= -i[\mathcal{H}_\text{fw}^s,\rho_\text{fw}] + \mathcal{L}[L_a]\rho_\text{fw} + \mathcal{L}[L_{bs}]\rho_\text{fw} +  \mathcal{\pounds}_\text{n}[L_{bs}]\rho_\text{fw} \;,
  \end{aligned}
 \end{equation}
 Here, we use the operator $L_{bs} = \sqrt{\kappa_b} b_{s_\circlearrowleft}$ and the noise-related Lindblad term
 \begin{equation}\label{eq:MEqnoise}
  \mathcal{\pounds}_\text{n}[L_{bs}]\rho_\text{fw} = N_p \mathcal{L}[L_{bs}]\rho_\text{fw} + N_p \mathcal{L}[L_{bs}^\dagger]\rho_\text{fw} - M_p \mathcal{L}^\prime [L_{bs}]\rho_\text{fw} - M_p^* \mathcal{L}^\prime [L_{bs}^\dagger]\rho_\text{fw} \;,
 \end{equation}
 where $N_p=\sinh^2(r_p)$, $M_p=e^{i\theta_p}\cosh(r_p)\sinh(r_p)$, and $\mathcal{L}^\prime[o]\rho = 2o\rho o - o o \rho -\rho o o$.

 In the backward-input case, the master equation of the system is given by
 \begin{equation}\label{eq:MEq2}
  \frac{\mathrm{d}}{\mathrm{d}t} \rho_\text{bw} = -i[\mathcal{H}_\text{bw},\rho_\text{bw}] + \mathcal{L}[L_a]\rho_\text{bw} + \mathcal{L}[L_b]\rho_\text{bw} \;.
 \end{equation}

 Note that the term $\mathcal{\pounds}_\text{n}[L_{bs}]\rho_\text{fw}$ in Eq.~\eqref{eq:MEq1s} is induced by quantum squeezing for the forward-input case. This noise-related term is absent in Eq.~\eqref{eq:MEq2} in the backward-input case.

 According to the input-output relation~\cite{pra.31.1985}, we have
 \begin{subequations}\label{eq:io}
  \begin{align}
  a_\text{out} &=a_{\text{in}} - \sqrt{2\kappa_{\text{ex1}}} a \;, \quad
   a_\text{out}^\dagger a_\text{out}  = a_\text{in}^\dagger a_\text{in} -\sqrt{2\kappa_\text{ex1}} \left(a_\text{in}^\dagger a +a^\dagger a_\text{in} \right) + 2\kappa_\text{ex1} a^\dagger a  \;,\\
 b_\text{out} &= \sqrt{2\kappa_\text{ex2}} b \;, \quad
  b_\text{out} ^\dagger b_\text{out}  = 2\kappa_\text{ex2}  b^\dagger b  \;,
  \end{align}
 \end{subequations}
 where $\kappa_\text{ex2}$ is the external decay rate of $\text{R}_\text{B}$.

 The transmissions are defined as
  \begin{equation}\label{eq:transmission}
  T_{12/21} = \frac{\langle a_\text{out}^\dagger a_\text{out}\rangle} {\langle a_\text{in}^\dagger a_\text{in}\rangle } \;,\quad
  T_{23} = \frac{\langle b_\text{out}^\dagger b_\text{out}\rangle }{\langle a_\text{in}^\dagger a_\text{in}\rangle }\;,
 \end{equation}
 where $T_{ij}$ is the transmission from port $i$ to port $j$, with $i,~j = 1,~2,~3$.
 According to Eqs.~\eqref{eq:io} and ~\eqref{eq:transmission}, replacing the operators $a_\text{in}$ and $a^\dag_\text{in}$ with their average value $\alpha_\text{in}$ and $\alpha_\text{in}^*$ and setting $\theta_p = 0$ (by adjusting the phase of the pump), we can solve the transmission for a coherent signal field by numerically solving Eq.~\eqref{eq:MEq1s} and~\eqref{eq:MEq2}.

\subsection{Elimination of squeezing-induced noise and single-photon input}\label{Sec:elimination_noise}
 In the forward-input case, to suppress the squeezing-induced noise, we input a broadband squeezed-vacuum field with squeezing parameter $r_e$ and reference phase $\theta_e$ into $\text{R}_\text{B}$ from port $3$.  The squeezed-vacuum field can be regarded as a squeezed-vacuum reservoir for CCW modes in $\text{R}_\text{B}$. Therefore, the dynamics of the system is described by the master equation~\cite{QOBook.1997}
 \begin{equation}\label{eq:MEq1sv}
  \begin{aligned}
    \frac{\mathrm{d}}{\mathrm{d}t}{\rho_\text{fw}^\text{sv}}=&-i[\mathcal{H}_\text{fw},\rho_\text{fw}^\text{sv}] + \mathcal{L}[L_a]\rho_\text{fw}^\text{sv} + (N_e+1)\mathcal{L}[L_b]\rho_\text{fw}^\text{sv}
     +N_e \mathcal{L}[L_b^\dagger]\rho_\text{fw}^\text{sv} - M_e \mathcal{L}^\prime[L_b]\rho_\text{fw}^\text{sv} - M_e^* \mathcal{L}^\prime[L_b^\dagger]\rho_\text{fw}^\text{sv} \;,
  \end{aligned}
 \end{equation}
 where $N_e=\sinh ^{2}\left(r_{e}\right)$ and $M_e= e^{-i \theta_{e}} \cosh \left(r_{e}\right) \sinh \left(r_{e}\right) $.

 Next, we respectively transform the Hamiltonian $\mathcal{H}_\text{fw}$ to $\mathcal{H}_\text{fw}^s$, and the operator $L_b$ to $L_b^s $ according to the Bogoliubov squeezing transformation   $b=\cosh(r_p)b_s-e^{-i\theta_p}\sinh(r_p)b_s^\dagger$. After the replacement, the master equation for the forward input can accordingly be written as
 \begin{equation}\label{eq:MEq1svs}
  \begin{aligned}
   \frac{\mathrm{d}}{\mathrm{d}t}{\rho_\text{fw}^\text{sv}}=&-i[\mathcal{H}_\text{fw}^s,\rho_\text{fw}^\text{sv}] + \mathcal{L}[L_a]\rho_\text{fw}^\text{sv} + (N_e+1)\mathcal{L}[L_b^s]\rho_\text{fw}^\text{sv}
   + N_e \mathcal{L}[L_b^{s\dagger}]\rho_\text{fw}^\text{sv} - M_e \mathcal{L}^\prime[L_b^s]\rho_\text{fw}^\text{sv} - M_e^* \mathcal{L}^\prime[L_b^{s\dagger}]\rho_\text{fw}^\text{sv}\\
   =&-i[\mathcal{H}_\text{fw}^s,\rho_\text{fw}^\text{sv}] + \mathcal{L}[L_a]\rho_\text{fw}^\text{sv} +  \mathcal{L}[L_{bs}]\rho_\text{fw}^\text{sv} + N_e^s \mathcal{L}[L_{bs}]\rho_\text{fw}^\text{sv}
   + N_e^s \mathcal{L} [L_{bs}^\dagger]\rho_\text{fw}^\text{sv} - M_e^s \mathcal{L}^\prime [L_{bs}]\rho_\text{fw}^\text{sv} - M_e^{s*} \mathcal{L}^\prime [L_{bs}^\dagger]\rho_\text{fw}^\text{sv} \;,
  \end{aligned}
 \end{equation}
 where
 \begin{subequations}
  \begin{align}
    N_e^s=& \cosh ^{2}\left(r_p\right) \sinh ^{2}\left(r_{e}\right)+\sinh ^{2}\left(r_p\right) \cosh ^{2}\left(r_{e}\right) +\frac{1}{2} \sinh \left(2 r_p\right) \sinh \left(2 r_{e}\right) \cos \left(\theta_{p}+\theta_{e}\right) \;, \\
    M_e^s=& \exp \left(i\theta_{p}\right) \left[\sinh \left(r_p\right) \cosh \left(r_{e}\right)+\exp \left[-i(\theta_{p}+\theta_{e})\right] \cosh \left(r_p\right) \sinh \left(r_{e}\right)\right]  \\
    &\times \left[\cosh \left(r_p\right) \cosh \left(r_{e}\right)+\exp \left[i(\theta_{p}+\theta_{e})\right] \sinh \left(r_{p}\right) \sinh \left(r_e\right)\right] \;.
  \end{align}
 \end{subequations}
 When $r_e=r_p$ and $\theta_e + \theta_p =\pm n\pi~(n=1,3,5,...)$, we have $N_e^s=0$ and $M_e^s=0$. Thus, Eq.~\eqref{eq:MEq1svs} becomes
 \begin{equation} \label{eq:MEq1svs0}
  \frac{\mathrm{d}}{\mathrm{d}t}{\rho_\text{fw}^\text{sv}}=-i[\mathcal{H}_\text{fw}^s,\rho_\text{fw}^\text{sv}] + \mathcal{L}[L_a]\rho_\text{fw}^\text{sv} + \mathcal{L}[L_{bs}]\rho_\text{fw}^\text{sv} \;.
 \end{equation}
 Here, the squezzing-noise-induced term $\mathcal{\pounds}_\text{n}[L_{bs}]\rho_\text{fw}$ in Eq.~\eqref{eq:MEq1s} is cancelled by the squeezed-vacuum field in the forward-input case. Therefore, the squeezed mode equivalently couples to a normal vacuum bath. As a result, the decay rate of the squeezed mode equals that of the original bare mode, i.e. $\kappa_{bs} = \kappa_b$. The term $\mathcal{L}[L_{bs}]\rho_\text{fw}$ with operator $L_{bs} = \sqrt{\kappa_b}b_{s_\circlearrowleft}$ describes the decay of the mode $b_{s_\circlearrowleft}$ with a rate $\kappa_b$.

 In the backward-input case, the squeezed-vacuum field from port $3$ has no influence on the system dynamics. So the motion of the system coupling to a normal vacuum bath is governed by the master equation
 \begin{equation} \label{eq:MEq2svs}
  \frac{\mathrm{d}}{\mathrm{d}t}{\rho_\text{bw}} = -i[\mathcal{H}_\text{bw},\rho_\text{bw}] + \mathcal{L}[L_a]\rho_\text{bw} + \mathcal{L}[L_b]\rho_\text{bw} \;.
 \end{equation}

 Then, we use a quantum cascaded system to simulate the propagation of single-photon pulses incident to ports $1$ and $2$ simultaneously~\cite{Gardiner1993,Carmichael1993,njp.14.2012}. In this quantum cascaded system, single-photon pulses emitted from the source resonator are input into our optical nonreciprocal device. Therefore, when the squeezing-vacuum field is applied, the master equation describing the quantum cascaded system for the forward-input case is given by
 \begin{equation}\label{eq:cascadedMEq1}
  \frac{\mathrm{d}}{\mathrm{d}t} \rho_\text{qcs,fw}^\text{sv} = -i[\mathcal{H}_d,\rho_\text{qcs,fw}^\text{sv}] - i[\mathcal{H}_\text{qcs,fw}^s,\rho_\text{qcs,fw}^\text{sv}] + \mathcal{L}[L_d]\rho_\text{qcs,fw}^\text{sv} + \mathcal{L}[L_a]\rho_\text{qcs,fw}^\text{sv} + \mathcal{L}[L_{bs}]\rho_\text{qcs,fw}^\text{sv} + \mathcal{\pounds}_\text{qcs,fw} \rho_\text{qcs,fw}^\text{sv} \;,
 \end{equation}
 where $\mathcal{H}_d=\Delta_d d^\dagger d$ is the Hamiltonian of the source resonator, $\Delta_d = \omega_d - \omega_\text{in}$, $\omega_d$ is the resonance frequency of the source resonator, and $\rho_\text{qcs,fw}^\text{sv}$ is the joint density matrix of the source resonator and our device in the forward-input case. Here we set $\Delta_d=0$.
 The Hamiltonian of the quantum cascaded system is $\mathcal{H}_\text{qcs,fw}^s = \Delta_a a_{_\circlearrowright}^\dagger a_{_\circlearrowright} + \Delta_b^s b_{s_\circlearrowleft}^{\dagger} b_{s_\circlearrowleft}+J_s(a_{_\circlearrowright}^\dagger b_{s_\circlearrowleft} + b_{s_\circlearrowleft}^{\dagger}a_{_\circlearrowright})$.
 The Lindblad operator $L_d=\sqrt{\kappa_\text{ex0}}d$ describes the external decay of the source resonator, where $\kappa_\text{ex0}$ is the decay rate from the source resonator to the device. To apply a Gaussian-like single-photon pulse, we set $\kappa_\text{ex0}(t) = \kappa_a \exp( - {(t - {\tau _\text{d}})^2}/2\tau _\text{p}^2)$, where $\tau_p$ is the pulse duration and $\tau_d$ is the pulse delay~\cite{pra.97.2018,prl.123.2019}. The relevant Lindblad terms are $\mathcal{L}[L_d]\rho = {\kappa_\text{ex0}}(2d\rho d^\dagger-d^\dagger d \rho-\rho d^\dagger d)$ and $\mathcal{\pounds}_\text{qcs,fw}\rho_\text{qcs,fw}^\text{sv} = \sqrt{4\kappa_\text{ex0}\kappa_\text{ex1}}\left( [a_{_\circlearrowright}^\dagger, d\rho] + [\rho d^\dagger, a_{_\circlearrowright}] \right)$.

 In the backward-input case, the master equation of the quantum cascaded system is given by
 \begin{equation}\label{eq:cascadedMEq2}
  \frac{\mathrm{d}}{\mathrm{d}t} \rho_\text{qcs,bw} = -i[\mathcal{H}_d,\rho_\text{qcs,bw}] - i[\mathcal{H}_\text{qcs,bw},\rho_\text{qcs,bw}] + \mathcal{L}[L_d]\rho_\text{qcs,bw} + \mathcal{L}[L_a]\rho_\text{qcs,bw} + \mathcal{L}[L_{b}]\rho_\text{qcs,bw} + \mathcal{\pounds}_\text{qcs,bw} \rho_\text{qcs,bw} \;,
 \end{equation}
 where $\mathcal{H}_\text{qcs,bw}= \Delta_a a_{_\circlearrowleft}^\dagger a_{_\circlearrowleft} +\Delta_b b_{_\circlearrowright}^\dagger b_{_\circlearrowright}+J(a_{_\circlearrowleft}^\dagger b_{_\circlearrowright} + b_{_\circlearrowright}^\dagger a_{_\circlearrowleft})$ and $\mathcal{\pounds}_\text{qcs,bw} \rho_\text{qcs,bw}^\text{sv} = \sqrt{4\kappa_\text{ex0}\kappa_\text{ex1}}\left( [a_{_\circlearrowleft}^\dagger, d\rho] + [\rho d^\dagger, a_{_\circlearrowleft}] \right)$, and $\rho_\text{qcs,bw}^\text{sv}$ is the joint density matrix of the source resonator and our device in the backward-input case.

\begin{figure}[ht]
  \centering
  \includegraphics[width=0.7\linewidth]{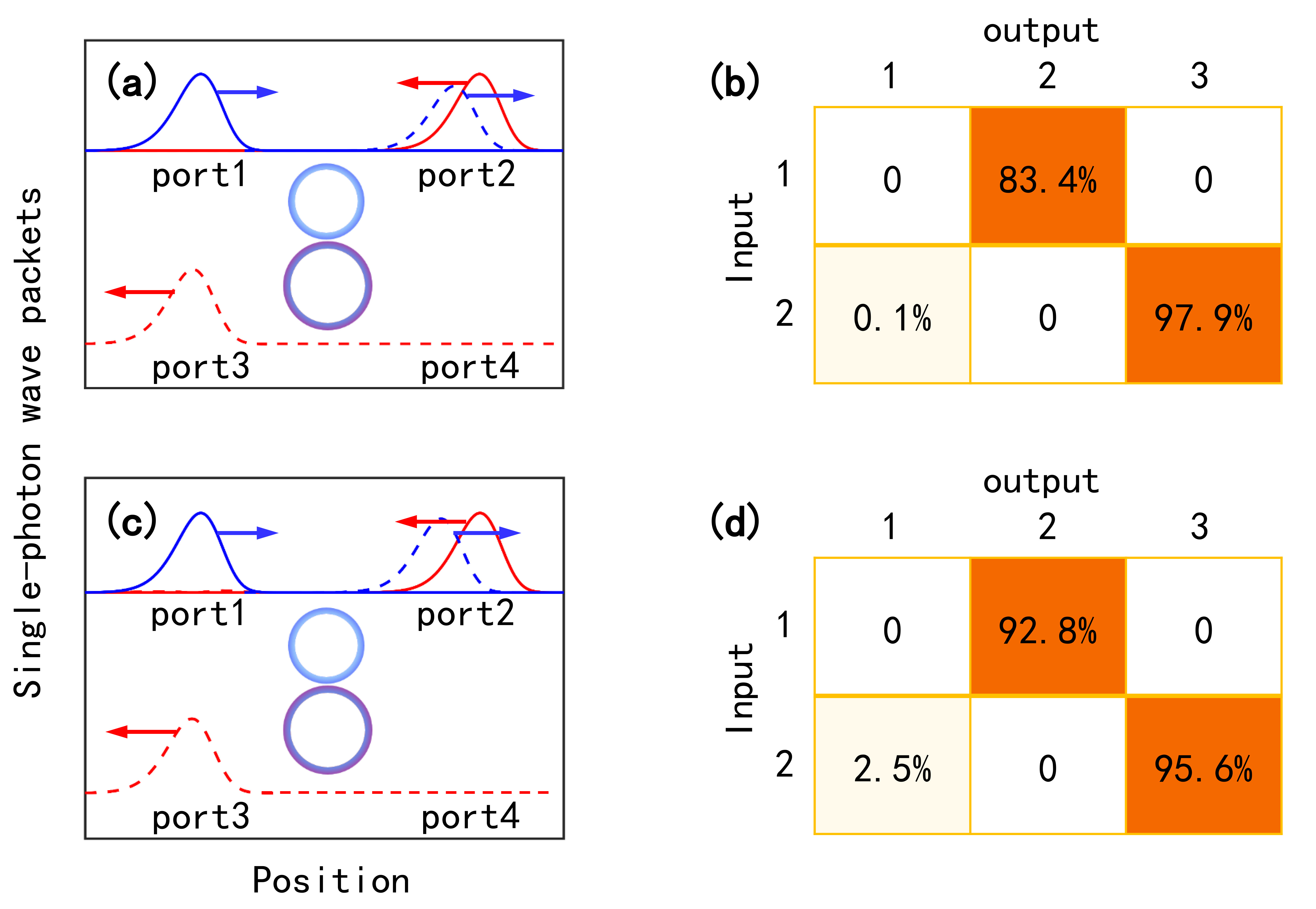} \\
  \caption{ (a) and (c) Propagation of single-photon pulses with a duration of $2\pi \times 6\kappa_a^{-1}$. The arrows indicate the propagating directions of the pulses. Blue (red) curves are for the incident (solid curves) and transmitted (dashed curves) pulses for the forward (backward) input. (b) and (d) Transmission matrix of the three-port quasi-circulator. (a) and (b) for the normal mode splitting (NMS) scenario with $\kappa_a = \kappa_b$, $\kappa_\text{ex1,2}/\kappa_a=0.99$, $J/\kappa_a=0.99$, $\Delta_p^b/\kappa_a=10.3$, $\Omega_p/\kappa_a=10$, and $\Delta_a=\Delta_b=0$. (c) and (d) for the mode resonance shift (MRS) scenario with $\kappa_a = \kappa_b$, $\kappa_\text{ex1,2}/\kappa_a=0.99$, $J/\kappa_a=2.8$, $\Delta_p^b/\kappa_a=15$, $\Omega_p/\kappa_a=13$, and $\Delta_a=\Delta_b=2.62\kappa_a$.}
  \label{fig:pulse}
 \end{figure}

 According to Eqs.~\eqref{eq:io} and ~\eqref{eq:transmission}, we can attain the propagation (left panel in Fig.~S\ref{fig:pulse}) and transmissions (right panel in Fig.~S\ref{fig:pulse}) of the single-photon pulses by solving Eq.~\eqref{eq:cascadedMEq1} and~\eqref{eq:cascadedMEq2} numerically. As show in Fig.~S\ref{fig:pulse}, the system can function as a three-port quasi-circulator, allowing single-photon pulses propagating along port $1 \rightarrow 2 \rightarrow 3$~\cite{adr.2.2021}.
 %
 \medskip

 \section{Quantum Langevin equation}
 According to Eq.~\eqref{eq:MEq1s}, the quantum Langevin equation for an arbitrary system operator $Q$ in the forward-input case is given by
 \begin{equation}\label{eq:qlEq0}
   \frac{\mathrm{d}}{\mathrm{d}t} Q = i[\mathcal{H}_\text{fw}^s,Q] + \mathcal{L}[L_a] Q +  \mathcal{L}[L_{bs}] Q + \mathcal{\pounds}_\text{n}[L_{bs}]Q \;,
 \end{equation}
 where $L_a = \sqrt{\kappa_a}a_{_\circlearrowright}$, $L_b = \sqrt{\kappa_b}b_{s_\circlearrowleft}$, $ \mathcal{\pounds}_\text{n}[L_{bs}]Q = N_p \mathcal{L}[L_{bs}]Q + N_p \mathcal{L}[L_{bs}^\dagger]Q - M_p \mathcal{L}^\prime [L_{bs}]Q - M_p^* \mathcal{L}^\prime [L_{bs}^\dagger]Q$, $\mathcal{L}[o]Q = 2o^\dagger Q o - Q o^\dagger o - o^\dagger o Q$, and $\mathcal{L}^\prime[o]Q = 2o Q o - Q o o - o o Q$. Hence, we have the equations of motion for the specific operators $Q = \{a_{_\circlearrowright}, ~b_{s_\circlearrowleft}, ~a_{_\circlearrowright}^{\dagger} b_{s_\circlearrowleft}, ~b_{s_\circlearrowleft}^{\dagger} b_{s_\circlearrowleft}, ~a_{_\circlearrowright}^{\dagger} a_{_\circlearrowright} \}$ reading as
 \begin{subequations} \label{eq:QLEq1}
  \begin{align}
    &\frac{\mathrm{d}}{\mathrm{d}t} a_{_\circlearrowright} = -(i\Delta_a + \kappa_a)a_{_\circlearrowright} + \sqrt{2\kappa_\text{ex1}}a_\text{in} - i J_s b_{s_\circlearrowleft}\;, \label{eq:QLEq1a}\\
    &\frac{\mathrm{d}}{\mathrm{d}t} b_{s_\circlearrowleft} = -(i\Delta_b^s+\kappa_b)b_{s_\circlearrowleft} - iJ_s a_{_\circlearrowright}\;, \label{eq:QLEq1b}\\
    &\frac{\mathrm{d}}{\mathrm{d}t} {a_{_\circlearrowright}^{\dagger} b_{s_\circlearrowleft}}  =
    [i\Delta_{ab}^s-\kappa_{ab}]a_{_\circlearrowright}^{\dagger} b_{s_\circlearrowleft} + \sqrt{2\kappa_\text{ex1}} a_\text{in}^\dagger b_{s_\circlearrowleft} - i J_s \Xi\;,\label{eq:QLEq1c}\\
    &\frac{\mathrm{d}}{\mathrm{d}t} {b_{s_\circlearrowleft}^{\dagger} b_{s_\circlearrowleft}}  =
    i J_s (a_{_\circlearrowright}^\dagger b_{s_\circlearrowleft} - a_{_\circlearrowright} b_{s_\circlearrowleft}^{\dagger}) - 2\kappa_b b_{s_\circlearrowleft}^{\dagger} b_{s_\circlearrowleft} +\Psi_\text{noise} \;, \label{eq:QLEq1d}\\
    & \frac{\mathrm{d}}{\mathrm{d}t} {a_{_\circlearrowright}^{\dagger} a_{_\circlearrowright}}  =
  - i J_s (a_{_\circlearrowright}^\dagger b_{s_\circlearrowleft} - a_{_\circlearrowright} b_{s_\circlearrowleft}^{\dagger}) + \sqrt{2\kappa_\text{ex1}}(a_\text{in}a_{_\circlearrowright}^{\dagger} + a_\text{in}^\dagger a_{_\circlearrowright}) - 2\kappa_a a_{_\circlearrowright}^{\dagger} a_{_\circlearrowright}\;. \label{eq:QLEq1e}
  \end{align}
 \end{subequations}
 Here, $\Psi_\text{noise} = 2\sinh^2(r_p)\kappa_b$, $\Delta_{ab}^s = \Delta_a-\Delta_b^s$, $\kappa_{ab}=\kappa_a+\kappa_b$ and $\Xi = a_{_\circlearrowright}^\dagger a_{_\circlearrowright} b_{s_\circlearrowleft} b_{s_\circlearrowleft}^{\dagger} - a_{_\circlearrowright} a_{_\circlearrowright}^\dagger b_{s_\circlearrowleft}^{\dagger} b_{s_\circlearrowleft}$. To a good approximation, using the commutation relations $[a,a^\dagger]=1$ and $[b_s,b_s^\dagger]=1$, we can derive $\Xi = a_{_\circlearrowright}^\dagger a_{_\circlearrowright} - b_{s_\circlearrowleft}^{\dagger} b_{s_\circlearrowleft}$. By setting $ \frac{\mathrm{d}}{\mathrm{d}t} {Q} = 0$, we can obtain the steady-state mean values of the operators
 \begin{subequations}\label{eq:ssoper1}
  \begin{align}
  &\left \langle a_{_\circlearrowright} \right \rangle_\text{ss} = \frac{(i\Delta_b^s+\kappa_b)\sqrt{2\kappa_\text{ex1}}\alpha_\text{in}}{(i\Delta_a+\kappa_a)(i\Delta_b^s+\kappa_b)+J_s^2}\;, \label{eq:ssoper1a}\\
  &\left \langle b_{s_\circlearrowleft} \right \rangle_\text{ss} = \frac{-iJ_s\sqrt{2\kappa_\text{ex1}}\alpha_\text{in}}{(i\Delta_a+\kappa_a)(i\Delta_b^s+\kappa_b)+J_s^2}\;, \label{eq:ssoper1b}\\
  &\left \langle a_{_\circlearrowright}^\dagger a_{_\circlearrowright}\right \rangle_\text{ss}  = \frac{2\kappa_\text{ex1}|\alpha_\text{in}|^2(\kappa_b^2+{\Delta_b^s}^2)}{\mathcal{G}_s} + \mathcal{N}_\text{noise} \;,\label{eq:ssoper1c}\\
  &\left \langle b_{s_\circlearrowleft}^{\dagger} b_{s_\circlearrowleft}\right \rangle_\text{ss}  = \frac{2\kappa_\text{ex1}|\alpha_\text{in}|^2J_s^2}{\mathcal{G}_s} + \varrho \mathcal{N}_\text{noise} \;.
  \end{align}
 \end{subequations}
 Here, $\mathcal{N}_\text{noise} = {\kappa_b(\kappa_a+\kappa_b) \sinh^2(r_p) J_s^2}\big/{\mathcal{Q}_s}$. The mean value $\alpha_\text{in} = \left \langle a_\text{in} \right \rangle$ is the coherent amplitude of the input signal field. We also have $\mathcal{G}_s=J_s^4+2J_s^2(\kappa_a \kappa_b - \Delta_a \Delta_b^s)+(\kappa_a^2+\Delta_a^2)(\kappa_b^2+{\Delta_b^s}^2)$, $\mathcal{Q}_s=J_s^2(\kappa_a+\kappa_b)^2+\kappa_a\kappa_b[(\kappa_a+\kappa_b)^2+{\Delta_{ab}^s}^2]$, and $\varrho = J_s^2(\kappa_a+\kappa_b)+\kappa_a[(\kappa_a+\kappa_b)^2+{\Delta_{ab}^s}^2]\big/(\kappa_a+\kappa_b)J_s^2$.

 According to Eqs.~\eqref{eq:transmission} and~\eqref{eq:ssoper1}, the steady-state transmission for port $1 \rightarrow 2$ is
 \begin{equation} \label{eq:T12}
  T_{12}=\frac{J_s^4+2 \zeta_s J_s^2+\Lambda_s}{\mathcal{G}_s} + \frac{2\kappa_\text{ex1}\mathcal{N}_\text{noise}}{|\alpha_\text{in}|^2}\;,
 \end{equation}
 where $\zeta_s=\kappa_a \kappa_b - 2\kappa_b \kappa_\text{ex1}-\Delta_a \Delta_b^s$ and $\Lambda_s=[(\kappa_a-2\kappa_\text{ex1})^2+\Delta_a^2](\kappa_b^2+{\Delta_b^s}^2)$.

 Similarly, according to Eq.~\eqref{eq:MEq2}, we obtain the quantum Langevin equation for an arbitrary system operator $Q$ in the back-input case
 \begin{equation}\label{eq:qlEq0}
   \frac{\mathrm{d}}{\mathrm{d}t} Q = i[\mathcal{H}_\text{bw},Q] + \mathcal{L}[L_a] Q +  \mathcal{L}[L_{b}] Q \;.
 \end{equation}
 Thus, we have the equations of motion for the specific operators $Q = \{a_{_\circlearrowleft}, ~b_{_\circlearrowright}, ~a_{_\circlearrowleft}^{\dagger} b_{_\circlearrowright}, ~b_{_\circlearrowright}^{\dagger} b_{_\circlearrowright}, ~a_{_\circlearrowleft}^{\dagger} a_{_\circlearrowleft} \}$:
 \begin{subequations} \label{eq:QLEq2}
  \begin{align}
   &\frac{\mathrm{d}}{\mathrm{d}t} a_{_\circlearrowleft} = -(i\Delta_a + \kappa_a)a_{_\circlearrowleft} + \sqrt{2\kappa_\text{ex1}}a_\text{in} - i J b_{_\circlearrowright}\;, \label{eq:QLEq2a}\\
   &\frac{\mathrm{d}}{\mathrm{d}t} b_{_\circlearrowright} = -(i\Delta_b+\kappa_b)b_{_\circlearrowright} - iJ a_{_\circlearrowleft}\;, \label{eq:QLEq2b}\\
   &\frac{\mathrm{d}}{\mathrm{d}t} {a_{_\circlearrowleft}^{\dagger} b_{_\circlearrowright}}  = (i\Delta_{ab}- \kappa_{ab})a_{_\circlearrowleft}^{\dagger} b_{_\circlearrowright} + \sqrt{2\kappa_\text{ex1}} a_\text{in}^\dagger b_{_\circlearrowright}  - i J \Xi \;, \label{eq:QLEq2c}\\
   &\frac{\mathrm{d}}{\mathrm{d}t} {b_{_\circlearrowright}^\dagger b_{_\circlearrowright}}  =
    i J (a_{_\circlearrowleft}^\dagger b_{_\circlearrowright} - a_{_\circlearrowleft} b_{_\circlearrowright}^\dagger) - 2\kappa_b b_{_\circlearrowright}^\dagger b_{_\circlearrowright}\;, \label{eq:QLEq2d}\\
   &\frac{\mathrm{d}}{\mathrm{d}t} {a_{_\circlearrowleft}^{\dagger} a_{_\circlearrowleft}}  =
   - i J (a_{_\circlearrowleft}^\dagger b_{_\circlearrowright} - a_{_\circlearrowright} b_{_\circlearrowleft}^\dagger) + \sqrt{2\kappa_\text{ex1}}(a_\text{in}a_{_\circlearrowleft}^{\dagger} + a_\text{in}^\dagger a_{_\circlearrowleft}) - 2\kappa_a a_{_\circlearrowleft}^{\dagger} a_{_\circlearrowleft}\;. \label{eq:QLEq2e}
  \end{align}
 \end{subequations}
 Here, we use the notation $\Delta_{ab} = \Delta_a-\Delta_b$, $\Xi = a_{_\circlearrowleft}^\dagger a_{_\circlearrowleft}  -  b_{_\circlearrowright}^{\dagger} b_{_\circlearrowright}$. Setting $ \frac{\mathrm{d}}{\mathrm{d}t} {Q} = 0$, we obtain the steady-state solutions
 \begin{subequations}\label{eq:ssoper2}
  \begin{align}
   &\left \langle a_{_\circlearrowleft} \right \rangle_\text{ss} = \frac{(i\Delta_b+\kappa_b)\sqrt{2\kappa_\text{ex1}}\alpha_\text{in}}{(i\Delta_a+\kappa_a)(i\Delta_b+\kappa_b)+J^2}\;, \label{eq:ssoper2a}\\
  &\left \langle b_{_\circlearrowright} \right \rangle_\text{ss} = \frac{-iJ\sqrt{2\kappa_\text{ex1}}\alpha_\text{in}}{(i\Delta_a+\kappa_a)(i\Delta_b+\kappa_b)+J^2}\;, \label{eq:ssoper2b}\\
  &\left\langle a_{_\circlearrowleft}^{\dagger} a_{_\circlearrowleft} \right \rangle_\text{ss} = \cfrac{2\kappa_\text{ex1}|\alpha_\text{in}|^2(\kappa_b^2+\Delta_b^2)}{\mathcal{G}}\;, \label{eq:ssoper2c}\\
  &\left\langle b_{_\circlearrowright}^\dagger b_{_\circlearrowright} \right \rangle_\text{ss} = \cfrac{2\kappa_\text{ex1}|\alpha_\text{in}|^2J^2}{\mathcal{G}}\;. \label{eq:ssoper2d}
  \end{align}
 \end{subequations}
  where $\mathcal{G} = J^4+2J^2(\kappa_a \kappa_b-\Delta_a \Delta_b)+(\kappa_a^2+\Delta_a^2)(\kappa_b^2+\Delta_b^2)$.
 According to Eqs.~\eqref{eq:transmission} and~\eqref{eq:ssoper2}, the steady-state transmissions in the backward-input case are given by
 \begin{subequations} \label{T213}
  \begin{align}
  &T_{21} = \cfrac{J^4+2 \zeta J^2+\Lambda}{\mathcal{G}} \;, \label{T213a} \\
  &T_{23} = \cfrac{4\kappa_\text{ex1}\kappa_\text{ex2}J^2}{\mathcal{G}} \;, \label{T213b}
  \end{align}
 \end{subequations}
 where $\zeta=\kappa_a \kappa_b - 2\kappa_b\kappa_\text{ex1} -\Delta_a \Delta_b$ and $\Lambda=[(\kappa_a-2\kappa_\text{ex1})^2+\Delta_a^2](\kappa_b^2+\Delta_b^2)$.

 After applying a phase-matched squeezed-vacuum field to drive $\text{R}_\text{B}$, as discussed in Sec.~\ref{Sec:elimination_noise}, the squeezing-induced noise can be completely eliminated. In this case, we have $\kappa_{bs} = \kappa_b$. Thus, the term $\mathcal{L}[L_{bs}]b_{s_\circlearrowleft}^{\dagger} b_{s_\circlearrowleft}$ with the operator $L_{bs} = \sqrt{\kappa_b}b_{s_\circlearrowleft}$ describes the decay of the mode $b_{s_\circlearrowleft}$ with a rate $\kappa_b$.
 Therefore, when the squeezed-vacuum field is applied, we can rewrite Eq.~\eqref{eq:QLEq1d} as
 \begin{equation} \label{eq:bb}
  \frac{\mathrm{d}}{\mathrm{d}t} { b_{s_\circlearrowleft}^{\dagger} b_{s_\circlearrowleft}}  =  i J_s \left(a_{_\circlearrowright}^\dagger b_{s_\circlearrowleft} - a_{_\circlearrowright} b_{s_\circlearrowleft}^{\dagger}\right) - 2\kappa_b b_{s_\circlearrowleft}^{\dagger} b_{s_\circlearrowleft}\;.
 \end{equation}
 Note that the noise term $\Psi_\text{noise}$ in Eq.~\eqref{eq:QLEq1d} is eliminated; but equations~\eqref{eq:QLEq1a}-\eqref{eq:QLEq1c} and~\eqref{eq:QLEq1e} have no change in this case. So, we can derive the steady-state mean value of the mode $a_{_\circlearrowright}$ which is given by
 \begin{equation}
  \left\langle a_{_\circlearrowright}^\dagger a_{_\circlearrowright} \right \rangle_\text{ss} = \cfrac{2\kappa_\text{ex1}|\alpha_\text{in}|^2(\kappa_b^2+{\Delta_b^s}^2)}{\mathcal{G}_s} \;.
 \end{equation}
 Here, the noise term $\mathcal{N}_\text{noise}$ in Eq.~\eqref{eq:ssoper1c} is also eliminated in this case.

 Therefore, the steady-state transmissions are obtain as
 \begin{equation} \label{eq:Tsv}
 T_{12}^\text{sv} = (J_s^4+2 \zeta_s J_s^2+\Lambda_s) \big/ \mathcal{G}_s \;, \quad T_{21}^\text{sv} = T_{21} \;, \quad
 T_{23}^\text{sv}=T_{23} \;.
 \end{equation}
 From Eq.~\eqref{eq:T12}, we can see that the noise-related term ${2\kappa_\text{ex1}\mathcal{N}_\text{noise}}\big/{|\alpha_\text{in}|^2}$ in the transmission $T_{12}$ can be completely eliminated by applying the squeezed-vacuum field.


 \section{Second order nonlinear parametric process}
 We now consider the full quantum description of the degenerate nonlinear parametric process in $\text{R}_\text{B}$. Then, the Hamiltonian for the forward-input case is given by (for simplicity, we replace $a_{_\circlearrowright}$ with $a$, $b_{s_\circlearrowleft}$ with $b$, and $c_{_\circlearrowleft}$ with $c$)
 \begin{equation}
  \begin{aligned}
    \mathcal{H}/\hbar = & \omega_a a^\dagger a + \omega_b b^\dagger b + \omega_c c^\dagger c  + J(a^\dagger b + b^\dagger a) + g \left( {b^\dagger}^2 c + b^2 c^\dagger \right) \\
     &+ i  \sqrt{2\kappa_\text{ex1}}\alpha_\text{in} \left(a^\dagger e^{-i\omega_\text{in} t} - a e^{i\omega_\text{in} t}\right) + i  \sqrt{2\kappa_{\text{ex2}}^p}\alpha_p \left(c^\dagger e^{-i\omega_p t} - c e^{i\omega_p t}\right) \;,
  \end{aligned}
 \end{equation}
 where $\omega_{a/b}$ is the resonance frequency of the fundamental signal mode in $\text{R}_\text{A}$ or $\text{R}_\text{B}$, $\omega_c$ is the frequency of the second-harmonic modes in $\text{R}_\text{B}$, $\alpha_\text{in} = \sqrt{2\pi P_\text{in} / \hbar\omega_\text{in}}$ is the coherent amplitude of the incident signal light with the power $P_\text{in}$, $\alpha_p = \sqrt{2\pi P_p / \hbar\omega_p}$ corresponds to the pump light with the power $P_p$ and the angular frequency $\omega_p$, and $\kappa_\text{ex2}^p$ is the external decay rate for the pump field mode in $\text{R}_\text{B}$. Note that the factor $2\pi$ in $\alpha_\text{in}$ and $\alpha_p$ is needed to keep the dimension consistent in the angular frequency.
 In the rotating frame defined by  $U= \exp[(-i \frac{\omega_p}{2}a^\dagger a -i \frac{\omega_p}{2} b^\dagger b - i \omega_p c^\dagger c)t]$, the Hamiltonian becomes
 \begin{equation}\label{eq:H1N1}
  \begin{aligned}
   \mathcal{H}/\hbar = & \Delta_p^a a^\dagger a + \Delta_p^b b^\dagger b + \Delta_p^c c^\dagger c + J(a^\dagger b + b^\dagger a) + g \left( {b^\dagger}^2 c + b^2 c^\dagger \right) \\
   &+ i  \sqrt{2\kappa_\text{ex1}}\alpha_\text{in} \left(a^\dagger e^{-i\Delta_\text{in} t} - a e^{i\Delta_\text{in} t}\right) + i  \sqrt{2\kappa_{\text{ex2}}^p}\alpha_p \left( c^\dagger - c \right) \;,
   \end{aligned}
 \end{equation}
 where $\Delta_p^{a/b} = \omega_{a/b} - \omega_p/2$, $\Delta_p^c = \omega_c - \omega_p$, and $\Delta_\text{in} = \omega_\text{in} - \omega_p/2$. The dynamical equation of $c$ can be solved by the Heisenberg equation
 \begin{equation}
   \dot{c} = i[\mathcal{H},c] - \kappa_p c = -(i\Delta_p^c + \kappa_p) c + \sqrt{2\kappa_{\text{ex2}}^p}\alpha_p - igb^2\;.
 \end{equation}
 Here, we consider a strong continuous pump field to excite the mode $c$ in $\text{R}_\text{B}$ with amplitude $\left \langle c \right\rangle \gg \left \langle b \right\rangle$. In this strong pump case, we can omit the last term and solve the steady state of mode $c$ as
 \begin{equation}
  \left \langle c \right\rangle_\text{ss} = \frac{\sqrt{2\kappa_{\text{ex2}}^p}\alpha_p}{i\Delta_p^c + \kappa_p} \;.
 \end{equation}
For a slowly varying mode $c$, we can replace $c$ with its steady-state mean value $\left \langle c \right\rangle_\text{ss}$ and obtain the Hamiltonian Eq.~\eqref{eq:H1N1} rewritten as
 \begin{equation}\label{eq:Hnonlinear}
  \mathcal{H}/\hbar =  \Delta_p^a a^\dagger a + \Delta_p^b b^\dagger b + J(a^\dagger b + b^\dagger a) + g  \left( {b^\dagger}^2 \left \langle c \right\rangle_\text{ss} + b^2 \left \langle c \right\rangle_\text{ss}^* \right) + i  \sqrt{2\kappa_\text{ex1}}\alpha_\text{in} \left(a^\dagger e^{-i\Delta_\text{in} t} - a e^{i\Delta_\text{in} t}\right) \;.
 \end{equation}
 Comparing Eq.~\eqref{eq:Hnonlinear} with Eq.~\eqref{eq:H1}, we can estimate the amplitude of the pump as
 \begin{equation}
   \Omega_p = 2 g|\left \langle c \right\rangle_\text{ss}| = 4g\sqrt{\frac{\pi\kappa_{\text{ex2}}^p P_p}{({\Delta_p^c}^2 + \kappa_p^2)\hbar \omega_p}} \;.
 \end{equation}
 The resonance pump field at frequency $\omega_p = \omega_c$ leads to $\Delta_p^c = 0$. The pump power is given by
 \begin{equation}
   P_p = \frac{\hbar \omega_p \kappa_p^2 \Omega_p^2} {16\pi g^2 \kappa_{\text{ex2}}^p} \;.
 \end{equation}
 To evaluate the optical transistor, we define the gain of the transistor as
 \begin{equation}
   G = \frac{P_\text{in}}{P_p}\Delta T = \frac{8\omega_\text{in} g^2\kappa_\text{ex2}^p \alpha_\text{in}^2}{\omega_p \kappa_p^2\Omega_p^2}\Delta T = \frac{2\kappa_\text{ex2}g^2\alpha_\text{in}^2}{\kappa_a^2\Omega_p^2}\Delta T \;.
 \end{equation}
 Here, owing to $\Delta_p^b \ll \{\omega_a,~\omega_b,~\omega_\text{in}, ~\omega_p$\}, we have applied the following approximations, $\omega_p=2\omega_\text{in}$, $\kappa_p = 2\kappa_a$, and $\kappa_\text{ex2}^p = 2\kappa_\text{ex2}$, to obtain the final form of the gain.

\medskip

%